\documentclass[%
reprint,
superscriptaddress,
 amsmath,amssymb,
 aps,
 pra,
]{revtex4-1}
\usepackage[dvipsnames]{xcolor}
\definecolor{mblue}{RGB}{42, 54, 144} 
\usepackage{hyperref}
\hypersetup{backref,pdfpagemode=FullScreen,colorlinks=true,breaklinks,urlcolor=mblue,linkcolor=mblue,citecolor=mblue}
\usepackage{graphicx}
\usepackage{dcolumn}
\usepackage{bm}
\usepackage{braket}
\usepackage{amssymb}
\usepackage{changes}
\usepackage{siunitx}

\begin{document}


\title{Double Degenerate Bose-Fermi Mixture of Strontium and Lithium}


\author{Zhu-Xiong Ye}
\affiliation{State Key Laboratory of Low-Dimensional Quantum Physics, Department of Physics, Tsinghua University, Beijing 100084, China}
\author{Li-Yang Xie}
\affiliation{State Key Laboratory of Low-Dimensional Quantum Physics, Department of Physics, Tsinghua University, Beijing 100084, China}

\author{Zhen Guo}
\affiliation{State Key Laboratory of Low-Dimensional Quantum Physics, Department of Physics, Tsinghua University, Beijing 100084, China}

\author{Xiao-Bin Ma}
\affiliation{State Key Laboratory of Low-Dimensional Quantum Physics, Department of Physics, Tsinghua University, Beijing 100084, China}

\author{Gao-Ren Wang}
\affiliation{School of Physics, Dalian University of Technology, Dalian 116024, China}

\author{Li You}
\affiliation{State Key Laboratory of Low-Dimensional Quantum Physics, Department of Physics, Tsinghua University, Beijing 100084, China}
\affiliation{Frontier Science Center for Quantum Information, Beijing, China}

\author{Meng Khoon Tey}
\email{mengkhoon\_tey@tsinghua.edu.cn}
\affiliation{State Key Laboratory of Low-Dimensional Quantum Physics, Department of Physics, Tsinghua University, Beijing 100084, China}
\affiliation{Frontier Science Center for Quantum Information, Beijing, China}


\date{\today}

\begin{abstract}
We report on the attainment of a degenerate Fermi gas of $\rm^{6}Li$ in contact with a Bose-Einstein condensate (BEC) of $^{84}$Sr. A degeneracy of $T/T_F=0.33(3)$ is observed with $1.6\times10^5$ $^{6}$Li atoms in the two lowest energy hyperfine states together with an almost pure BEC of $3.1\times10^5$ $^{84}$Sr atoms. The elastic s-wave scattering length between $^6$Li and $^{84}$Sr is estimated to be $|a_{\rm^{6}Li-\rm^{84}Sr}|=(7.1_{-1.7}^{+2.6})a_0$ ($a_0$ being the Bohr radius) from measured interspecies thermalization rates in an optical dipole trap.
\end{abstract}

\pacs{}

\maketitle


Ultracold gas mixtures composed of different species provide a useful platform to study few- and many-body physics, including BEC-BCS crossover \cite{Bloch2008,Diener2010}, Efimov state \cite{Levinsen2009,Barontini2009,ChinCheng2014} and crystaline quantum phase \cite{Petrov2007}. They are also ideal starting points to prepare heteronuclear ground-state molecules \cite{Deiglmayr2008,Sage2005} which hold great promises for the study of ultracold chemical reactions \cite{Ospelkaus2010,Hu2019}, quantum computation and simulation \cite{Rabl2006,DeMille2002} and precision measurement \cite{Carr2009}. Besides, large mass mismatch gas mixtures can also be used to study impurity in superfluids \cite{Targonskatargonska2010,Spiegelhalder2009}, mass imbalanced Cooper pairs \cite{Gezerlis2009,Trenkwalder2011,Wu2006,Iskin2008} and heteronuclear trimer states \cite{Pires2014,DIncao2006}. Previously, quantum degenerate mixtures of alkali-metal atoms have been the main powerhouse for such studies.

In recent years, mixtures of alkali-metal and alkaline-earth(-like) atoms have attracted much attention. Significant progresses have been made in both theories~\cite{Dulieu2010SrLipotential,gopakumar2013ab,Mitroy2003,Standard1985,Stanton1994,Kotochigova2011,Chen2014,Pototschnig2016,Augustovicova2012,Tomza2013,Guo2013,shao2017ground,Tohme2015,Allouche1994,kajita2012,Brue2012,MahmoudKoreka2018SrX} and experiments~\cite{Hara2011,Hansen2011,Vaidya2015,Pasquiou2013,Guttridge2018,Aoki2013,Bruni2016,Schwanke2017,kraft2009Ca,WolfgangE2014DroplrtRbSr,JackSimons1998LiCa,JVerges1994BaLi}. In particular, Feshbach resonances in RbSr~\cite{Barbe2018} and LiYb~\cite{green2019feshbach} have been observed, paving ways to produce ground-state group-I+group-II molecules. One of the motivations to produce such molecules is that they possess a magnetic dipole moment on top of an electric dipole moment. This property makes such molecules good candidates for studies of lattice-spin models \cite{Micheli2006}, collective spin excitations in optical lattice~\cite{perez2010} and precision measurements of, for examples, the electric-dipole moment (EDM) of the electron~\cite{Meyer2009} and proton-to-electron mass ratio~\cite{kajita2012}. So far, mixtures of alkali-metal and alkaline-earth(-like) atoms which have been cooled down to quantum degeneracy include Li-Yb \cite{Hara2011,Hansen2011}, Rb-Yb \cite{Vaidya2015}, Rb-Sr \cite{Pasquiou2013} and Cs-Yb \cite{Guttridge2018}.

In this work, we report the production of the first quantum degenerate mixture composed of alkali-metal $ \rm^{6}Li $ and alkaline-earth-metal $ \rm^{84}Sr $ in a far-off-resonant optical dipole trap. Previously, we have estimated the $s$-wave scattering length between $^{88}$Sr and $^6$Li atoms to be $|a_\mathrm{^6Li-^{88}Sr}|=(14.4^{+4.9}_{-3.2})a_0$ from measuring the interspecies thermalization rates~\cite{Ma2019,*[Erratum:]Ma2019Erratum}. Earlier experience seems to disfavor this number as too small to support efficient sympathetic cooling for typical conditions near quantum degeneracy, which typically requires good elastic $s$-wave scattering length of the order of 100$a_0$ instead. Moreover, changing the $^{88}$Sr to other isotopes is not expected to change the $^6$Li-Sr scattering substantially. This is because $^6$Li is much lighter than strontium, changing $^{88}$Sr to $^{84}$Sr, for example, changes the $^6$Li-$^x$Sr reduced mass by merely 0.3\%. Since the ground-state molecular potential of Li-Sr can only support roughly 23 vibrational bound states~\cite{Dulieu2010SrLipotential,gopakumar2013ab}, a 0.3\% variation in the reduced mass would hardly have any impact on the overall energy structure of the Li-Sr molecules, and thus the interspecies scattering length. Our simulation based on the potential from \cite{Dulieu2010SrLipotential} predicts a change of only a few $a_0$ in the $^6$Li-Sr scattering length from $^{88}$Sr to $^{84}$Sr. Indeed, based on the same measurement method detailed in Ref. ~\cite{Ma2019,*[Erratum:]Ma2019Erratum}, we estimate $|a_\mathrm{^6Li-^{84}Sr}|$ to be $(7.1_{-1.7}^{+2.6})a_0$.

Even though changing to the $^{84}$Sr isotope does not result in a more favorable interspecies thermalization, it is more useful for our present goal of realizing a double degenerate mixture of Li and Sr. This is because $^{84}$Sr, which has a $s$-wave scattering length of $a_\mathrm{^{84}Sr-^{84}Sr}=123a_0$, can be readily cooled by evaporation to quantum degeneracy by itself~\cite{Stellmer2009,Killian2009SrBEC}, in contrast to $^{88}$Sr with $a_\mathrm{^{88}Sr-^{88}Sr}=-2a_0$. On the other hand, due to the lack of efficient sympathetic cooling,  the $^6$Li gas, in a mixture of its two lowest hyperfine sublevels, must also be able to support rapid thermalization by itself for efficient evaporative cooling. As the $^6$Li atoms in the two lowest hyperfine sublevels do not interact at zero magnetic field, we choose to perform evaporation of the $^{84}$Sr-$^6$Li mixture at a magnetic field of 330\,G. At this field, the $s$-wave scattering length between the two lowest hyperfine sub-states of $^6$Li is $a_{12}=-290a_0$. This field is, however, not expected to have much effects on the Sr-Sr and Sr-Li scattering lengths since the ground state of bosonic Sr has no magnetic dipole moment and thus does not couple with various spins of $^6$Li unless higher order couplings are included~\cite{Barbe2018,Brue2012}.

In short, our strategy towards a double degenerate Li and Sr mixture is to employ a Sr isotope and a magnetic field that would allow efficient evaporative cooling of individual species. At the end of evaporation, we obtain Fermi degenerate $^{6}$Li with $ 8.2\times10^{4}$ atoms in each spin at $T/T_F = 0.33(3)$, in contact with an almost pure Bose-Einstein condensate (BEC) of $^{84}$Sr with $3.1\times10^{5}$ atoms.

This article is organized as follows: Section~\ref{sec:loadingSrLiInODT} describes the main experiment setup and how we sequentially load $^6$Li, and then $^{84}$Sr into a 1064-nm optical dipole trap (ODT). Section~\ref{sec:evaporation_stage} discusses the evaporation cooling procedure as well as the evidences for the double degenerate Bose-Fermi mixture. The properties of the mixture over the evaporation are analyzed and discussed in Sec.~\ref{sec:propertiesOfMixture}. Section~\ref{sec:conclusion} concludes the article.

\section{Preparing $^{84}$Sr and $^6$Li atoms in optical dipole trap}\label{sec:loadingSrLiInODT}

\begin{figure*}
\centering	
		\includegraphics[width=1.5\columnwidth]{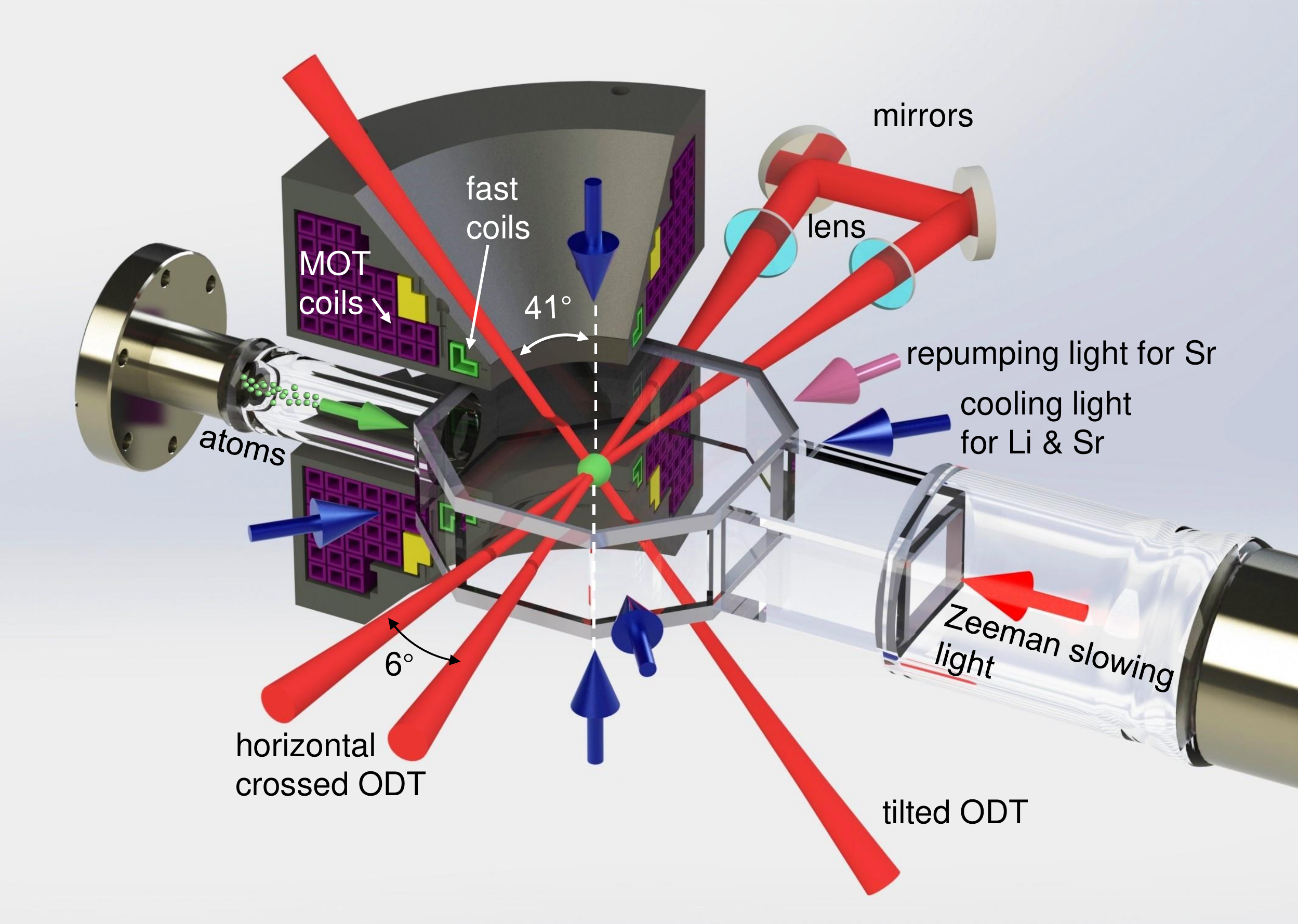}
		\caption{Illustration of the experimental setup. The main science chamber is an octagonal glass cell. Overlapped cooling light beams for both species are represented by the blue arrows~\cite{coolingbeams}. The $^6$Li+$^{84}$Sr atomic beams (green arrow) are overlapped and co-propagating, so as their corresponding Zeeman slowing light beams (red arrow). The MOT coils and the ``fast'' coils can operate in either Helmholtz or anti-Helmholtz configurations to provide bias magnetic fields and magnetic field gradients. The far-off resonant 1064-nm crossed optical dipole trap (CODT) for both species is formed at the common crossing of two horizontal beams and a beam tilted at $41^{\circ}$ to the vertical axis~\cite{dipoletraplasers}. }
\label{FIG:SETUP}
\end{figure*}

Figure \ref{FIG:SETUP} illustrates the heart of our setup, which has been described in detail in Ref.\,\cite{Ma2019,*[Erratum:]Ma2019Erratum}. In brief, overlapped light beams for cooling both lithium and strontium are indicated by blue arrows~\cite{coolingbeams}. Copropagating $^6$Li+$^{84}$Sr atomic beams and their corresponding Zeeman slowing beams are represented by a green arrow and a red arrow, respectively. The far-off resonant crossed optical trap (CODT) for both species is constructed by crossing three 1064-nm light beams (red beams) at their respective waists: two horizontal beams crossing at $6^{\circ}$ have a common waist of 32\,\si{\micro m}, the other is tilted at $41^{\circ}$ to the vertical axis and has a waist of 54\,\si{\micro m}~\cite{dipoletraplasers}.

\begin{figure*}
\centering
		\includegraphics[width=1.6\columnwidth]{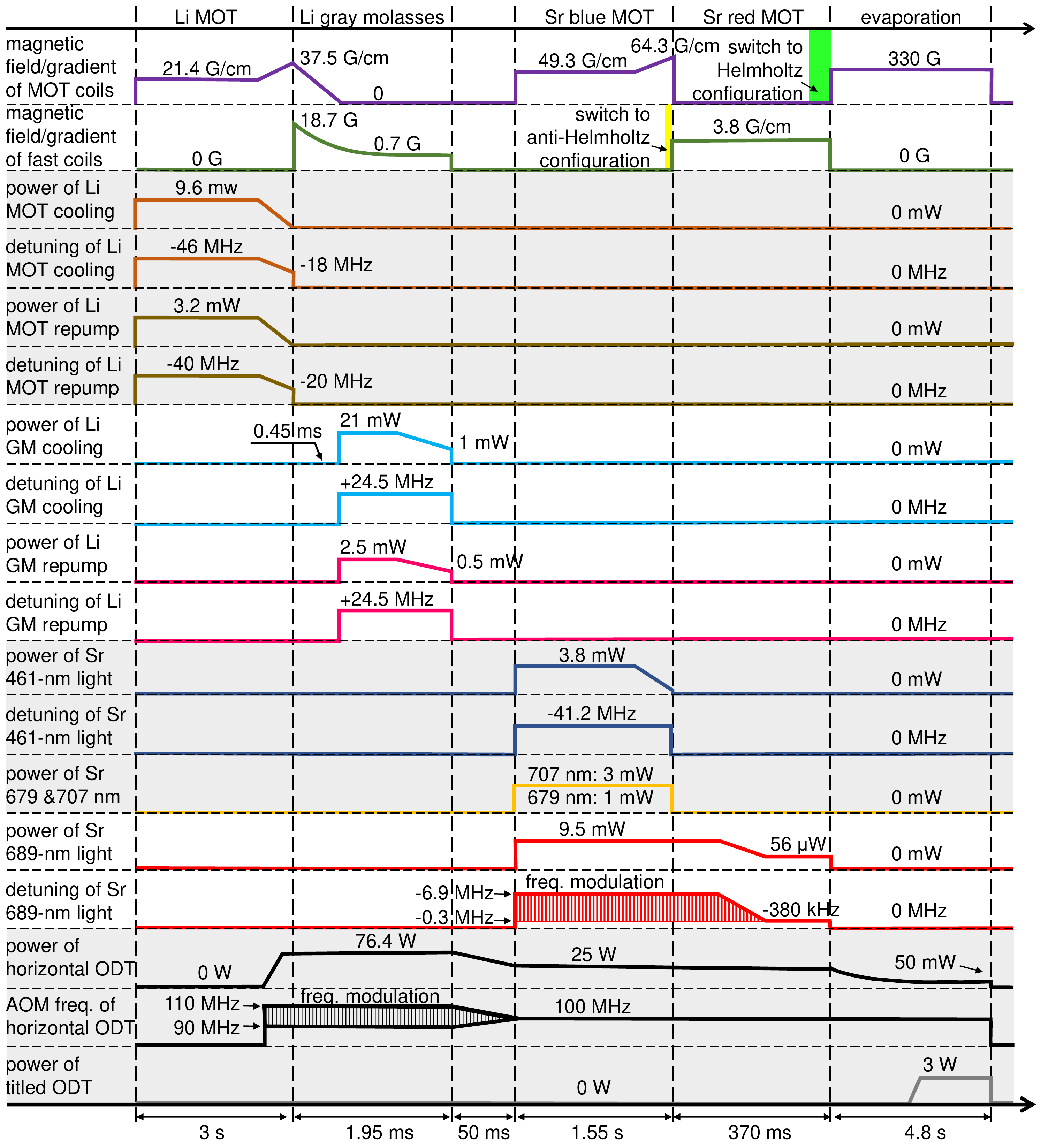}
		\caption{Schematic for main controls and timings for our experiment procedure. The reported powers of the various MOT beams are those of a single beam. Each MOT beam has a waist of $\sim$6\,mm. }
		\label{FIG:SEQUENCE}
\end{figure*}

As the magnetic field gradients for optimal loading and cooling of the lithium and strontium differ significantly, we sequentially (1) load the $^{6}$Li MOT, perform gray-molasses cooling and then transfer $^6$Li atoms into the horizontal crossed optical dipole trap (HCODT); (2) load the $^{84}$Sr 461-nm ``blue'' MOT, cool $^{84}$Sr atoms using the 689-nm ``red'' MOT and transfer them into the same HCODT; (3) evaporatively cool both $^6$Li and $^{84}$Sr into quantum degeneracy at a bias magnetic field of 330\,G by reducing the trap depth. The sequence of our experiment is detailed in Fig.\,\ref{FIG:SEQUENCE}.

In more detail, the experiment starts with magneto-optical trapping of $^6$Li atoms using cooling and repumping light locked to the 671-nm $^{2} {S}_{1 / 2} \rightarrow {^{2} {P}_{3 / 2}} $ $(\mathrm{D} 2)$ transitions and a magnetic field gradient of 21.4 G/cm. A cloud of $ 1.6\times10^{8} $ atoms is accumulated with a temperature $\sim3$ $\rm mK$ after 3 seconds. The MOT is then compressed (CMOT) by ramping the magnetic field gradient to 37.5 G/cm (and changing the powers and detuning of the cooling and repumping light simultaneously, see Fig.\,\ref{FIG:SEQUENCE}) in 3\,ms. This process reduces the gas temperature to $\sim690$\,\si{\micro K}. To further cool and increase the phase space density of the atomic cloud, we apply the gray molasses (GM) cooling technique \cite{Burchianti2014}. The GM can only work efficiently when the magnetic field is smaller than 0.1\,G.  Due to the induction of the MOT coils, however, the residual magnetic field at the cloud takes 1.1\,ms to drop below 0.1\,G. But lithium atom is relatively light, a 1.1-ms free expansion of the lithium cloud would greatly reduce the loading efficiency into the HCODT. As a remedy, a second set of coils, called ``fast coils'', in a Helmholtz configuration, dynamically compensate for the residual magnetic field from the MOT coils (see Fig.~\ref{FIG:SEQUENCE}), reducing the free expansion time from 1.1 ms to 0.45 ms. After 1.5\,ms of GM cooling, a cloud with $1\times10^{8}$ atoms at $\sim50$\,\si{\micro K} is obtained. As the $2S_{1/2}|F=1/2\rangle \rightarrow 2P_{1/2}|F'=3/2\rangle $ repumping light is switched off 0.1\,ms earlier than $2S_{1/2}|F=3/2\rangle \rightarrow 2P_{1/2}|F'=3/2\rangle $ cooling light at the end of the GM, atoms are equally distributed in the $|F=1/2, m_F=\pm 1/2\rangle$ states.

Transfer of the $^6$Li atoms into the HCODT (without the 41$^\circ$-tilted beam) happens concurrently with the GM process. To improve loading efficiency, the trapping volume of the HCODT is enlarged by $ 4\sim5 $ times at the beginning of the GM. This is done by modulating the frequency of the radio frequency (rf) applied to the acousto-optic modulator (AOM) which diffracts the horizontal ODT beam, from 90\,MHz to 110\,MHz, at a modulating frequency of 2.6\,MHz. The modulation amplitude is reduced linearly to 0 in 50 \,ms after the GM process (Fig.~\ref{FIG:SEQUENCE}). Meanwhile, the power of the HCODT beam is reduced linearly from 76.4\,W to 25\,W, stopping at a HCODT depth of $ k_B\times 1.9$\,mK ($k_B$ being the Boltzmann constant) for $^6$Li. At the end of the GM stage, about $ 3.3\times{10}^6 $ atoms are loaded into the HCODT at a temperature of $ \sim $300\,\si{\micro K}.

While the $^6$Li atoms are trapped in the HCODT, Zeeman slowed $^{84}$Sr atoms are cooled and captured in a ``blue'' MOT operating on a 461-nm light, red detuned from the $5s^2~{^1S_0}-5s5p~^1P_1$ transition  by 41.2\,MHz, at a magnetic field gradient of 49.3 G/cm. As weak leak from the cooling cycle populates the $5s5p~{{^{3}P}_2}$ metastable state at this stage~\cite{Stellmer2009}, two repumping light beams at 679\,nm and 707\,nm (pink arrow in Fig.~\ref{FIG:SETUP}) are shined onto the blue MOT to bring these atoms back into the cooling cycle. After 1.5-s loading, the $^{84}$Sr atom cloud is compressed by increasing the magnetic field gradient to 64.3\,G/cm within 50\,ms. Meanwhile, the power of each 461-nm beam is decreased from 3.8\,mW to 0.2\,mW.  At the end of the blue MOT stage, about $4.5\times10^7$ $^{84}$Sr atoms at a temperature of $\sim2$\,mK are collected.

Further cooling of the $^{84}$Sr atoms is achieved in a ``red'' MOT operating on the 7.5-kHz-linewidth $^1S_0-^3P_1$ transition at 689 nm, at a magnetic field gradient of 3.8 G/cm. To increase the capturing velocity of the red MOT, we frequency modulate the light, producing a detuning ranging from -6.9 MHz to -300 kHz with a spacing of 50 kHz. After a 75-ms modulation cooling period (which makes sure that almost all atoms are captured from the blue MOT into the red MOT), the cloud is compressed by ramping down the frequency modulation within 145\,ms, ending with a single-frequency MOT at a detuning of -380\,kHz (see Fig.~\ref{FIG:SEQUENCE}). At the same time, the power of each 689-nm beam is reduced from 9.5\,mW to 56\,\si{\micro W}. The single frequency MOT is then operated for another 150\,ms to further cool down the atoms. More than $ 2.1\times10^7$ atoms at a temperature 3.4\,\si{\micro K} are obtained at the end of the red MOT stage.

Loading of the $^{84}$Sr atoms into the HCODT occurs at the single-frequency red MOT stage. In order to achieve efficient loading, we optimize the overlap between the HCODT and the single-frequency red MOT by adjusting the bias magnetic fields. During the whole Sr-MOT stages, the power of HCODT is held constant at 25\,W, corresponding to a trap depth of $k_B\times 1.7$\,mK for strontium. Note that operation of the Sr blue MOT would result in loss of the $^6$Li atoms in the HCODT, presumably due to light-assisted collision loss~\cite{Gallagher1989}. We adjust the loading times for each species to obtain a favorable ratio in the atom numbers for subsequent evaporative cooling. With a 1.5-s blue MOT loading, we obtain $1.2\times{10}^6 $ $^6$Li atoms and $1.2\times{10}^7 $ $^{84}$Sr atoms in the HCODT at a common temperature of $\sim$ 180\,\si{\micro K} after thermalization for 500 ms between the two species.

\section{Evaporation to a double degenerate mixture}\label{sec:evaporation_stage}

For an optical trap from 1064-nm light, the ratios of trap depth and trap frequency for $^6$Li vs $^{84}$Sr are $U_\mathrm{^6Li}/U_\mathrm{^{84}Sr} = 1.1 $ and $\omega_\mathrm{^6Li}/\omega_\mathrm{^{84}Sr} = 3.9$, respectively. The almost equal depth for Li and Sr is useful for keeping the temperatures of Li and Sr nearly the same in the early stage of the evaporation despite weak interspecies interactions, since the temperature of a gas is typically about 1/10 of the $U/k_B$ after adequate evaporation (see Ref.~\cite{OHara2001} and Sec.~\ref{sec:propertiesOfMixture}).

Forced evaporation of the mixture starts by reducing the power of HCODT roughly exponentially from 25\,W to 50\,mW in 4.8\,s. After 1.5\,s into the forced evaporation, the power of the 41$^\circ$-tilted ODT is raised linearly from 0 to 3\,W in 450\,ms and then held constant till the end of the evaporation (see Fig.\,\ref{FIG:SEQUENCE}). At the end of the evaporation, we obtain a Fermi-degenerate $^{6}$Li gas with $ 8.2\times10^{4}$ atoms in each spin state at $T/T_F = 0.33(3)$, in contact with an almost pure BEC of $^{84}$Sr with $3.1\times10^{5}$ atoms. At this moment, the CODT frequencies and depth are, respectively, $(\omega_x,\omega_y,\omega_z)=2\pi\times(718,720,150)$\,Hz and $k_{B}\times5.6$\,$\si{\micro K}$ for Li, and $2\pi\times(144,183,37)$\,Hz and $k_{B}\times0.35$\,$\si{\micro K}$ for $^{84}$Sr, taking the gravitation sag into consideration.

\begin{figure}
\includegraphics[width=\columnwidth]{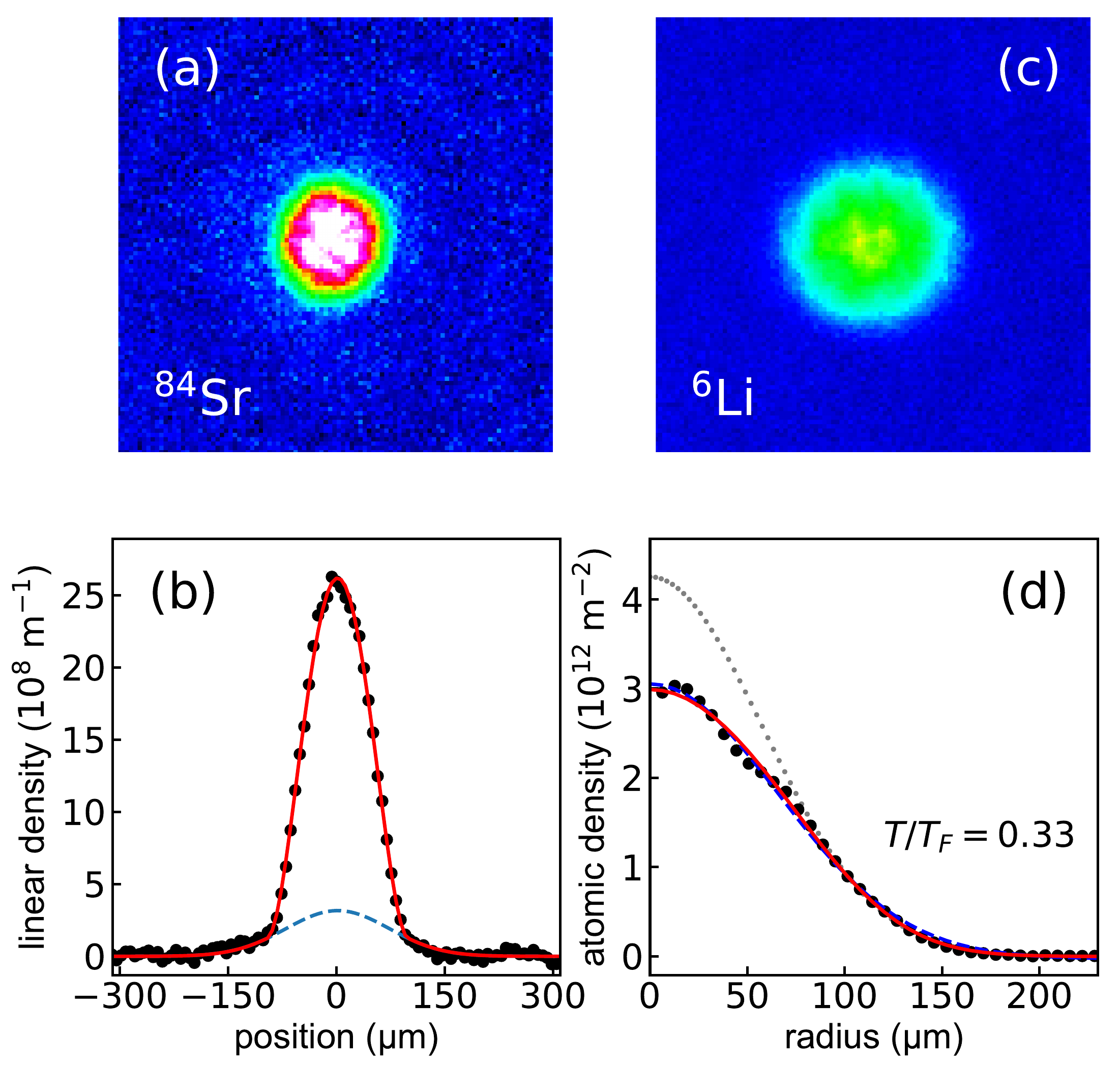}
\caption{Evidence of quantum degenerate Bose $^{84}$Sr and Fermi $^{6}$Li gases. (a) Absorption image of a BEC of $^{84}$Sr, 22\,ms after releasing from the trap. (b) Integrated density profiles of (a). The red solid line represents a fit with a bimodal distribution, while the blue dashed line denotes a Gaussian fit to the thermal part. (c) Absorption image of a quantum degenerate $^6$Li Fermi gas, 2\,ms after releasing from the trap. (d) The azimuthally averaged density distribution of the $^6$Li Fermi gas, averaged over 10 measurements. A fit using Eq.~(\ref{eq:FermiDiracDist}) (solid red line) deviates slightly from a Gaussian (dashed blue line). If we fit only the outer thermal wing (grey dotted line), outside the disk with a radius $\sqrt{2}w$ (where $w$ is the $1/e$ width of a Gaussian fit to the full distribution), the deviation becomes even more evident.}
\label{BECandFermiGas}
\end{figure}

As evidence for the double degeneracy, we show in Fig.\,\ref{BECandFermiGas} the absorption images and density profiles of the $^{84}$Sr BEC and the Fermi-degenerate $^{6}$Li gas after free expansion. The 1D integrated density profile of the $^{84}$Sr gas after a 22-ms time-of-flight [Fig.\,\ref{BECandFermiGas}(b)] exhibits a textbook-like bimodal distribution, signifying a BEC in contact with a small thermal component at a temperature of 105(10)\,nK. To quantify the degeneracy of the $^6$Li Fermi gas, we determine the ratio of temperature $T$ to the Fermi temperature $T_F$ by fitting the azimulthally averaged density profile of the $^6$Li gas after a 2-ms time-of-flight [Fig.\,\ref{BECandFermiGas}(d)] using~\cite{integratedFermiDirac}
\begin{equation}\label{eq:FermiDiracDist}
n(r)= A {\mathrm{Li}_{2}\left(-\zeta e^{-\frac{\rho^{2}}{2 \sigma^{2}}}\right)}.
\end{equation}
Here, Li$_n$ represents the $n$th-order polylogarithm function, and $\rho$ is the radius from the center of the cloud. $A$, $\sigma$, $\zeta$ are fitting parameters. The width $\sigma$ and fugacity $\zeta$ of the cloud can be used to extract $T=m \sigma^{2} / k_{B} t_\mathrm{tof}^{2}$ ($t_\mathrm{tof}$ being the time-of-flight) and $T/T_F=[\rm -6Li_3(-\zeta)]^{-1/3}$, respectively. The former gives degeneracy depth $T/T_F$, given that $T_{F}=\hbar \bar{\omega}(6N_{ \uparrow})^{\frac{1}{3}}/{k_{B}}$ [$\bar{\omega}=(\omega_x\omega_y\omega_z)^{1/3}$, $N_\uparrow$ is number of atom in one of the $^6$Li spins]. The latter is used for consistency check~\cite{Tey2010doubledegenerate} for results from the former when $\zeta>2$ (corresponding to $T/T_F<0.46$). For Fig.\,\ref{BECandFermiGas}(d), we obtain from the former method $T=539(31)$\,nK, $T_F=1.62$(10)\,$\si{\micro K}$, thus $T/T_F=0.33(3)$, which is consistent with $T/T_F=0.31(4)$ from the latter method.

\section{Properties of the mixture over the evaporation}\label{sec:propertiesOfMixture}

In this section, we analyze properties of the mixture at different moments of the evaporation process. The temperatures and atom numbers of the gases are obtained from absorption images, taken at 330 G, after free-expansion. The trap properties are computed from the powers and beam widths of the trapping light beams.

\begin{figure}
		\includegraphics[width=\columnwidth]{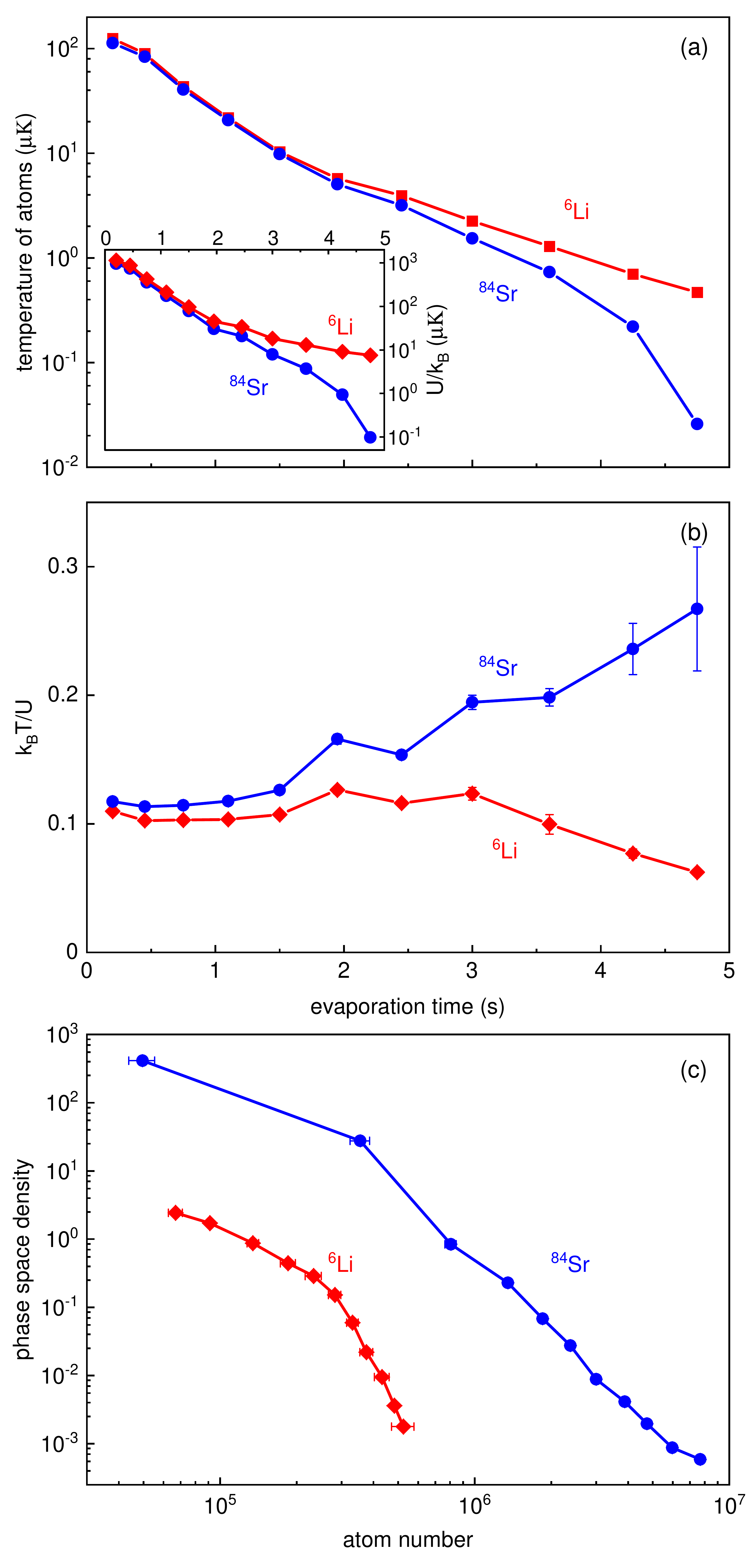}
		\caption{Temperatures $T$, trap depths $U$, and phase space densities at various stages of the evaporation process. (a) The temperatures of the $^6$Li and $^{84}$Sr gases show strong correlations with the respective depths of the ODT (inset), signifying insufficient interspecies thermalization towards the end of the evaporation. The error bars from 5 measurements are smaller than the symbols in this log scale plot. (b) Ratios of average thermal energy $ k_BT $ to the trap depth $ U $. (c) Phase space densities versus atom numbers. The PSD is defined as $n\lambda_T^3$, where $n$ is the gas density at the center of the trap, $\lambda_T=\sqrt{2\pi\hbar^2/m k_B T}$ being the thermal de Broglie wavelength. Atom number of $^6$Li represents that of a single component.}
\label{fig:evaporationcurve}
\end{figure}

Figure~\ref{fig:evaporationcurve}(a) plots the temperatures of the $^{84}$Sr and $^6$Li gases as a function of evaporation time. It shows that the temperatures of the two species are equal in the first 2\,s of the evaporation, but deviates substantially towards the end of the process. This suggests that heat exchange between the two species is efficient initially, but diminishes at the later stage of the evaporation. Without sufficient interspecies thermalization, the temperature of each gas is mainly affected by their corresponding trap depths $U$~\cite{OHara2001}. This is evident by noticing the similarities in the curves for the temperatures (main figure) and the computed trap depths $U$ (inset). As $U$ is reduced by the gravity more strongly for the heavier $^{84}$Sr than for the lighter $^6$Li, the $^{84}$Sr gas gets much colder than the $^6$Li one at the end of the evaporation. Fig.~\ref{fig:evaporationcurve}(b) shows the ratios $k_B T/U$ for the respective gases over the evaporation time. For $^6$Li, $k_B T/U$ stays roughly at the level of 0.1 throughout the evaporation, getting slightly smaller over time. For $^{84}$Sr, this parameter increases from approximately 0.1 in the beginning to 0.26 in the end. Figure~\ref{fig:evaporationcurve}(c) displays phase-space densities (PSDs) of the gases versus atom numbers. Roughly speaking, both the $^{84}$Sr and $^6$Li gases gain about 3 orders of magnitude in the PSDs after losing 1 order of magnitude in the atom number, signifying rather efficient cooling. The PSD of $^6$Li nevertheless levels off towards the end of the evaporation due to Fermi pressure of the gas. It should be noted that, at the end of the evaporation, the gravity sags the center of the $^6$Li($^{84}$Sr) cloud by $\sim0.5$\,$\si{\micro m}$($8.6$\,$\si{\micro m}$), while the Thomas-Fermi radius (assuming zero temperature) of $^6$Li($^{84}$Sr) are $\sim14.8$\,$\si{\micro m}$($5.9$ $\si{\micro m}$). This means that the two clouds remain fully overlaped till the very end of the evaporation process.

\begin{figure}
	\begin{center}
		\includegraphics[width=\columnwidth]{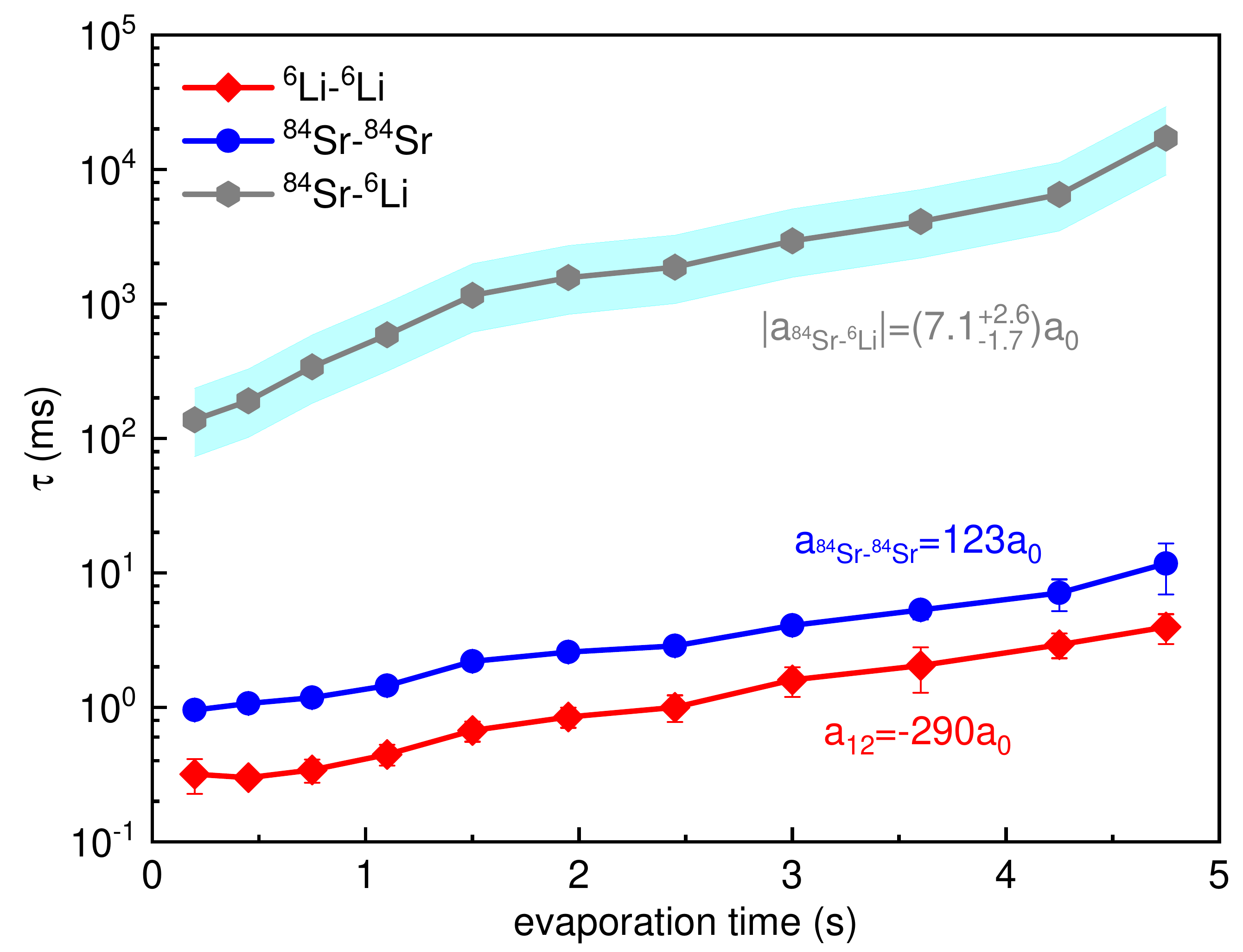}
		\caption{The computed intraspecies and interspecies thermal equilibrium time constants $\tau$ of $^6$Li and $^{84}$Sr atoms during evaporation. The blue shaded region represents uncertainty arising from the uncertainty in the interspecies $s$-wave scattering length. Influence of quantum degeneracy to collision cross sections, which would have effects to the last few points, are ignored in this calculation.}
\label{fig:ThermalEquilibriumRate}
	\end{center}
\end{figure}

In Fig.~\ref{fig:ThermalEquilibriumRate}, we show the interspecies and intraspecies thermal equilibrium time constants $\tau$, defined as $\tau=-\frac{1}{\Delta T}\frac{d(\Delta T)}{dt}$ ($\Delta T$ being the temperature difference between two thermal components)~\cite{Ma2019,*[Erratum:]Ma2019Erratum,thermalizationTimeConst}, as a function of the evaporation time. The figure shows that the intraspecies thermalization time constants of $^{84}$Sr and $^6$Li remain below 10\,ms throughout the evaporation. Such a value is small enough to guarantee good intraspecies thermalization over the whole evaporation process, which lasts over a few seconds. On the other hand, the interspecies thermalization is two to three orders of magnitude slower due to a much smaller interspecies scattering cross section $\propto a_s^2$ ($a_s$ being $s$-wave scattering length). The interspecies thermalization time constant increases from 100\,ms in the beginning to 10\,s at the end of the evaporation. This supports the observation in Fig.~\ref{fig:evaporationcurve} where the temperature difference between the two species grows over the evaporation process.

\begin{figure}
	\begin{center}
		\includegraphics[width=\columnwidth]{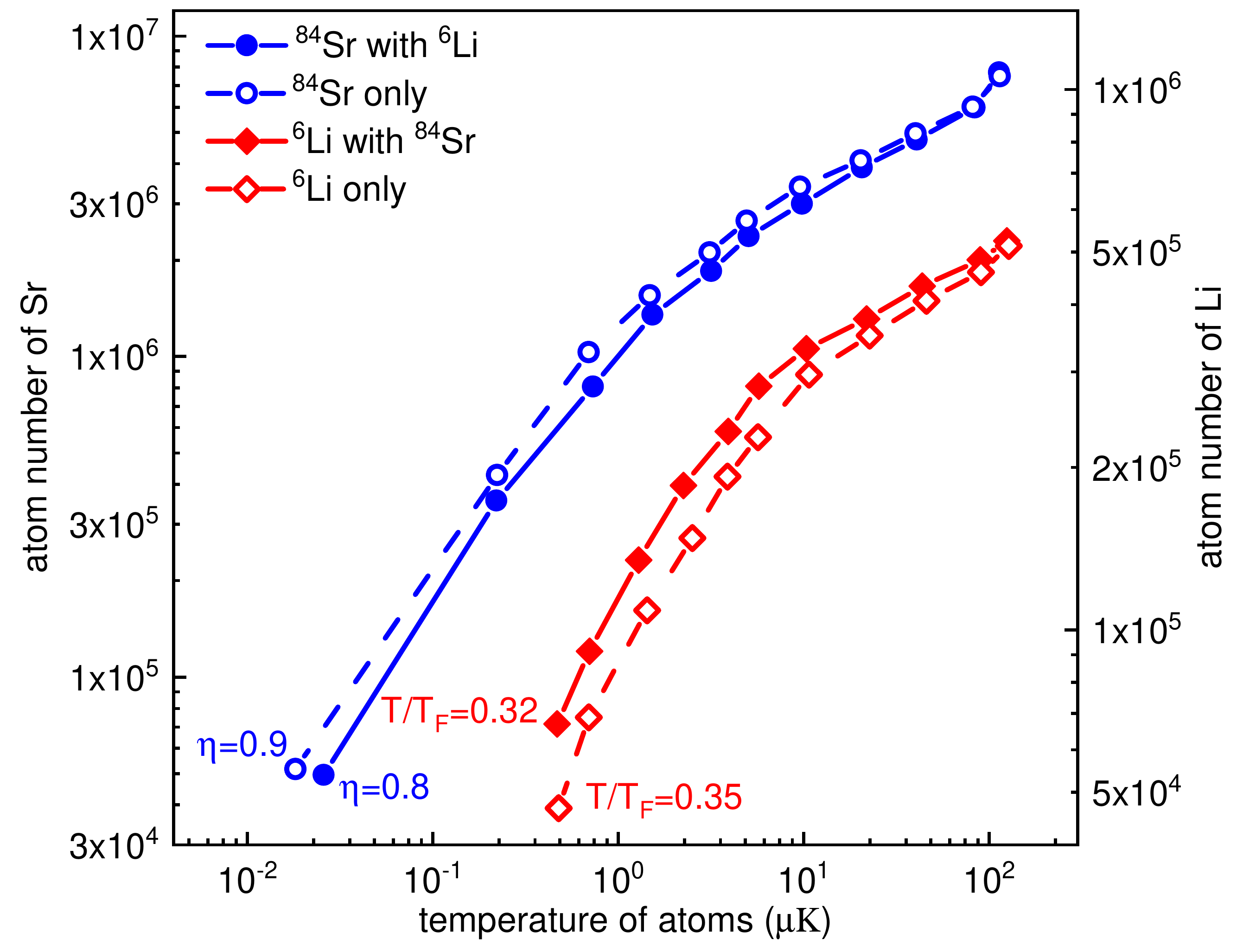}
		\caption{Effects of sympathetic cooling. The presence of $^{84}$Sr helps bringing the $^6$Li Fermi gas into deeper degeneracy. $ \eta=N_c/N$ represents the degeneracy of $^{84}$Sr, $N_c$ being the atom numbers in the condensate and $N$ is the total atom number of the gas. Error bars are smaller than the symbols. Atom number of $^6$Li represents that of a single component. }
\label{fig:effectsOfSympatheticCooling}
	\end{center}
\end{figure}

To see if the weak interspecies interactions play any role at all during the evaporative cooling process, we repeat the experiment without $^{84}$Sr or without $^6$Li. For each scenario, we keep the initial numbers of $^6$Li or $^{84}$Sr atoms at the beginning of the evaporation the same as those of the mixture experiment. The results, which are shown in Fig.~\ref{fig:effectsOfSympatheticCooling}, clearly demonstrate that the presence of $^{84}$Sr helps bringing the $^6$Li Fermi gas into deeper degeneracy. The latter is achieved by increasing the number of $^6$Li atoms at the expense of $^{84}$Sr atoms for any given ODT power, a clear signature of sympathetic cooling. This phenomenon can be readily understood given that the trap depth of $^{84}$Sr is always lower than that of $^6$Li.

\section{conclusion}\label{sec:conclusion}
In summary, we realize the first quantum degenerate mixture of fermionic $\rm^{6}Li$ and bosonic $\rm^{84}Sr$ atoms in a crossed optical dipole trap, and determine the elastic $s$-wave scattering length between $\rm^{6}Li$ and $\rm^{84}Sr$ atoms to be $|a_{\rm^{6}Li-\rm^{84}Sr}|=(7.1_{-1.7}^{+2.6})a_0$ by measuring interspecies thermalization rates. Our results pave the way to studies including Li-Sr polar molecules and impurity superfluids. Further improvement for our experiment includes adding a gradient magnetic field to lower the trap depth for the $^6$Li at the end of the evaporation process, so as to further reduce(increase) the temperature(quantum degeneracy) of the gas (we are not able to further reduce the power of the trapping light since this would result in spilling of the heavier $^{84}$Sr at the moment).

We note in passing that we have also realized double degenerate mixture of $^6$Li and $^{84}$Sr at zero magnetic field where there is no intraspecies $^6$Li collisions. Here, the $^6$Li gas can only be sympathetically cooled by the $^{84}$Sr atoms. This process causes much more atom loss in both species compared to evaporation reported above at 330\,G. We realize a degenerate $^6$Li Fermi gas at $T/T_F = 0.50(7)$ with total atom number of $N_{\rm Li}=4.9\times10^{4}$, coexisting with a $^{84}$Sr BEC of $ 2.1\times10^{5}$ atoms in this case. In the future, we plan to probe the binding energy spectrum of near dissociation Li-Sr molecules to determine the long-range dispersion coefficients of the ground-state molecular potential. Such information would be useful for producing ground-state molecules of Li-Sr.

We thank Xibo Zhang for helpful discussions. This work is supported by National Key Research and Development Program of China under Grant Nos 2018YFA0306503, and the National Natural Science Foundation of China under Grant Nos 91636213, 91736311, 11574177, 91836302 and 11654001, and Open Research Fund Program of the State Key Laboratory of Low-Dimensional Quantum Physics (KF201802).



\bibliographystyle{apsrev4-1}

\begin{thebibliography}{68}%
\makeatletter
\providecommand \@ifxundefined [1]{%
 \@ifx{#1\undefined}
}%
\providecommand \@ifnum [1]{%
 \ifnum #1\expandafter \@firstoftwo
 \else \expandafter \@secondoftwo
 \fi
}%
\providecommand \@ifx [1]{%
 \ifx #1\expandafter \@firstoftwo
 \else \expandafter \@secondoftwo
 \fi
}%
\providecommand \natexlab [1]{#1}%
\providecommand \enquote  [1]{``#1''}%
\providecommand \bibnamefont  [1]{#1}%
\providecommand \bibfnamefont [1]{#1}%
\providecommand \citenamefont [1]{#1}%
\providecommand \href@noop [0]{\@secondoftwo}%
\providecommand \href [0]{\begingroup \@sanitize@url \@href}%
\providecommand \@href[1]{\@@startlink{#1}\@@href}%
\providecommand \@@href[1]{\endgroup#1\@@endlink}%
\providecommand \@sanitize@url [0]{\catcode `\\12\catcode `\$12\catcode
  `\&12\catcode `\#12\catcode `\^12\catcode `\_12\catcode `\%12\relax}%
\providecommand \@@startlink[1]{}%
\providecommand \@@endlink[0]{}%
\providecommand \url  [0]{\begingroup\@sanitize@url \@url }%
\providecommand \@url [1]{\endgroup\@href {#1}{\urlprefix }}%
\providecommand \urlprefix  [0]{URL }%
\providecommand \Eprint [0]{\href }%
\providecommand \doibase [0]{http://dx.doi.org/}%
\providecommand \selectlanguage [0]{\@gobble}%
\providecommand \bibinfo  [0]{\@secondoftwo}%
\providecommand \bibfield  [0]{\@secondoftwo}%
\providecommand \translation [1]{[#1]}%
\providecommand \BibitemOpen [0]{}%
\providecommand \bibitemStop [0]{}%
\providecommand \bibitemNoStop [0]{.\EOS\space}%
\providecommand \EOS [0]{\spacefactor3000\relax}%
\providecommand \BibitemShut  [1]{\csname bibitem#1\endcsname}%
\let\auto@bib@innerbib\@empty
\bibitem [{\citenamefont {Bloch}\ \emph {et~al.}(2008)\citenamefont {Bloch},
  \citenamefont {Dalibard},\ and\ \citenamefont {Zwerger}}]{Bloch2008}%
  \BibitemOpen
  \bibfield  {author} {\bibinfo {author} {\bibfnamefont {I.}~\bibnamefont
  {Bloch}}, \bibinfo {author} {\bibfnamefont {J.}~\bibnamefont {Dalibard}}, \
  and\ \bibinfo {author} {\bibfnamefont {W.}~\bibnamefont {Zwerger}},\ }\href
  {\doibase 10.1103/RevModPhys.80.885} {\bibfield  {journal} {\bibinfo
  {journal} {Rev. Mod. Phys.}\ }\textbf {\bibinfo {volume} {80}},\ \bibinfo
  {pages} {885} (\bibinfo {year} {2008})}\BibitemShut {NoStop}%
\bibitem [{\citenamefont {Diener}\ and\ \citenamefont
  {Randeria}(2010)}]{Diener2010}%
  \BibitemOpen
  \bibfield  {author} {\bibinfo {author} {\bibfnamefont {R.~B.}\ \bibnamefont
  {Diener}}\ and\ \bibinfo {author} {\bibfnamefont {M.}~\bibnamefont
  {Randeria}},\ }\href {\doibase 10.1103/PhysRevA.81.033608} {\bibfield
  {journal} {\bibinfo  {journal} {Phys. Rev. A}\ }\textbf {\bibinfo {volume}
  {81}},\ \bibinfo {pages} {033608} (\bibinfo {year} {2010})}\BibitemShut
  {NoStop}%
\bibitem [{\citenamefont {Levinsen}\ \emph {et~al.}(2009)\citenamefont
  {Levinsen}, \citenamefont {Tiecke}, \citenamefont {Walraven},\ and\
  \citenamefont {Petrov}}]{Levinsen2009}%
  \BibitemOpen
  \bibfield  {author} {\bibinfo {author} {\bibfnamefont {J.}~\bibnamefont
  {Levinsen}}, \bibinfo {author} {\bibfnamefont {T.~G.}\ \bibnamefont
  {Tiecke}}, \bibinfo {author} {\bibfnamefont {J.~T.~M.}\ \bibnamefont
  {Walraven}}, \ and\ \bibinfo {author} {\bibfnamefont {D.~S.}\ \bibnamefont
  {Petrov}},\ }\href {\doibase 10.1103/PhysRevLett.103.153202} {\bibfield
  {journal} {\bibinfo  {journal} {Phys. Rev. Lett.}\ }\textbf {\bibinfo
  {volume} {103}},\ \bibinfo {pages} {153202} (\bibinfo {year}
  {2009})}\BibitemShut {NoStop}%
\bibitem [{\citenamefont {Barontini}\ \emph {et~al.}(2009)\citenamefont
  {Barontini}, \citenamefont {Weber}, \citenamefont {Rabatti}, \citenamefont
  {Catani}, \citenamefont {Thalhammer}, \citenamefont {Inguscio},\ and\
  \citenamefont {Minardi}}]{Barontini2009}%
  \BibitemOpen
  \bibfield  {author} {\bibinfo {author} {\bibfnamefont {G.}~\bibnamefont
  {Barontini}}, \bibinfo {author} {\bibfnamefont {C.}~\bibnamefont {Weber}},
  \bibinfo {author} {\bibfnamefont {F.}~\bibnamefont {Rabatti}}, \bibinfo
  {author} {\bibfnamefont {J.}~\bibnamefont {Catani}}, \bibinfo {author}
  {\bibfnamefont {G.}~\bibnamefont {Thalhammer}}, \bibinfo {author}
  {\bibfnamefont {M.}~\bibnamefont {Inguscio}}, \ and\ \bibinfo {author}
  {\bibfnamefont {F.}~\bibnamefont {Minardi}},\ }\href {\doibase
  10.1103/PhysRevLett.103.043201} {\bibfield  {journal} {\bibinfo  {journal}
  {Phys. Rev. Lett.}\ }\textbf {\bibinfo {volume} {103}},\ \bibinfo {pages}
  {043201} (\bibinfo {year} {2009})}\BibitemShut {NoStop}%
\bibitem [{\citenamefont {Tung}\ \emph {et~al.}(2014)\citenamefont {Tung},
  \citenamefont {Jim\'enez-Garc\'{\i}a}, \citenamefont {Johansen},
  \citenamefont {Parker},\ and\ \citenamefont {Chin}}]{ChinCheng2014}%
  \BibitemOpen
  \bibfield  {author} {\bibinfo {author} {\bibfnamefont {S.-K.}\ \bibnamefont
  {Tung}}, \bibinfo {author} {\bibfnamefont {K.}~\bibnamefont
  {Jim\'enez-Garc\'{\i}a}}, \bibinfo {author} {\bibfnamefont {J.}~\bibnamefont
  {Johansen}}, \bibinfo {author} {\bibfnamefont {C.~V.}\ \bibnamefont
  {Parker}}, \ and\ \bibinfo {author} {\bibfnamefont {C.}~\bibnamefont
  {Chin}},\ }\href {\doibase 10.1103/PhysRevLett.113.240402} {\bibfield
  {journal} {\bibinfo  {journal} {Phys. Rev. Lett.}\ }\textbf {\bibinfo
  {volume} {113}},\ \bibinfo {pages} {240402} (\bibinfo {year}
  {2014})}\BibitemShut {NoStop}%
\bibitem [{\citenamefont {Petrov}\ \emph {et~al.}(2007)\citenamefont {Petrov},
  \citenamefont {Astrakharchik}, \citenamefont {Papoular}, \citenamefont
  {Salomon},\ and\ \citenamefont {Shlyapnikov}}]{Petrov2007}%
  \BibitemOpen
  \bibfield  {author} {\bibinfo {author} {\bibfnamefont {D.~S.}\ \bibnamefont
  {Petrov}}, \bibinfo {author} {\bibfnamefont {G.~E.}\ \bibnamefont
  {Astrakharchik}}, \bibinfo {author} {\bibfnamefont {D.~J.}\ \bibnamefont
  {Papoular}}, \bibinfo {author} {\bibfnamefont {C.}~\bibnamefont {Salomon}}, \
  and\ \bibinfo {author} {\bibfnamefont {G.~V.}\ \bibnamefont {Shlyapnikov}},\
  }\href {\doibase 10.1103/PhysRevLett.99.130407} {\bibfield  {journal}
  {\bibinfo  {journal} {Phys. Rev. Lett.}\ }\textbf {\bibinfo {volume} {99}},\
  \bibinfo {pages} {130407} (\bibinfo {year} {2007})}\BibitemShut {NoStop}%
\bibitem [{\citenamefont {Deiglmayr}\ \emph {et~al.}(2008)\citenamefont
  {Deiglmayr}, \citenamefont {Grochola}, \citenamefont {Repp}, \citenamefont
  {M{\"{o}}rtlbauer}, \citenamefont {Gl{\"{u}}ck}, \citenamefont {Lange},
  \citenamefont {Dulieu}, \citenamefont {Wester},\ and\ \citenamefont
  {Weidem{\"{u}}ller}}]{Deiglmayr2008}%
  \BibitemOpen
  \bibfield  {author} {\bibinfo {author} {\bibfnamefont {J.}~\bibnamefont
  {Deiglmayr}}, \bibinfo {author} {\bibfnamefont {A.}~\bibnamefont {Grochola}},
  \bibinfo {author} {\bibfnamefont {M.}~\bibnamefont {Repp}}, \bibinfo {author}
  {\bibfnamefont {K.}~\bibnamefont {M{\"{o}}rtlbauer}}, \bibinfo {author}
  {\bibfnamefont {C.}~\bibnamefont {Gl{\"{u}}ck}}, \bibinfo {author}
  {\bibfnamefont {J.}~\bibnamefont {Lange}}, \bibinfo {author} {\bibfnamefont
  {O.}~\bibnamefont {Dulieu}}, \bibinfo {author} {\bibfnamefont
  {R.}~\bibnamefont {Wester}}, \ and\ \bibinfo {author} {\bibfnamefont
  {M.}~\bibnamefont {Weidem{\"{u}}ller}},\ }\href {\doibase
  10.1103/PhysRevLett.101.133004} {\bibfield  {journal} {\bibinfo  {journal}
  {Phys. Rev. Lett.}\ }\textbf {\bibinfo {volume} {101}},\ \bibinfo {pages}
  {133004} (\bibinfo {year} {2008})}\BibitemShut {NoStop}%
\bibitem [{\citenamefont {Sage}\ \emph {et~al.}(2005)\citenamefont {Sage},
  \citenamefont {Sainis}, \citenamefont {Bergeman},\ and\ \citenamefont
  {DeMille}}]{Sage2005}%
  \BibitemOpen
  \bibfield  {author} {\bibinfo {author} {\bibfnamefont {J.~M.}\ \bibnamefont
  {Sage}}, \bibinfo {author} {\bibfnamefont {S.}~\bibnamefont {Sainis}},
  \bibinfo {author} {\bibfnamefont {T.}~\bibnamefont {Bergeman}}, \ and\
  \bibinfo {author} {\bibfnamefont {D.}~\bibnamefont {DeMille}},\ }\href
  {\doibase 10.1103/PhysRevLett.94.203001} {\bibfield  {journal} {\bibinfo
  {journal} {Phys. Rev. Lett.}\ }\textbf {\bibinfo {volume} {94}},\ \bibinfo
  {pages} {203001} (\bibinfo {year} {2005})}\BibitemShut {NoStop}%
\bibitem [{\citenamefont {Ospelkaus}\ \emph {et~al.}(2010)\citenamefont
  {Ospelkaus}, \citenamefont {Ni}, \citenamefont {Wang}, \citenamefont {{De
  Miranda}}, \citenamefont {Neyenhuis}, \citenamefont {Qu{\'{e}}m{\'{e}}ner},
  \citenamefont {Julienne}, \citenamefont {Bohn}, \citenamefont {Jin},\ and\
  \citenamefont {Ye}}]{Ospelkaus2010}%
  \BibitemOpen
  \bibfield  {author} {\bibinfo {author} {\bibfnamefont {S.}~\bibnamefont
  {Ospelkaus}}, \bibinfo {author} {\bibfnamefont {K.~K.}\ \bibnamefont {Ni}},
  \bibinfo {author} {\bibfnamefont {D.}~\bibnamefont {Wang}}, \bibinfo {author}
  {\bibfnamefont {M.~H.}\ \bibnamefont {{De Miranda}}}, \bibinfo {author}
  {\bibfnamefont {B.}~\bibnamefont {Neyenhuis}}, \bibinfo {author}
  {\bibfnamefont {G.}~\bibnamefont {Qu{\'{e}}m{\'{e}}ner}}, \bibinfo {author}
  {\bibfnamefont {P.~S.}\ \bibnamefont {Julienne}}, \bibinfo {author}
  {\bibfnamefont {J.~L.}\ \bibnamefont {Bohn}}, \bibinfo {author}
  {\bibfnamefont {D.~S.}\ \bibnamefont {Jin}}, \ and\ \bibinfo {author}
  {\bibfnamefont {J.}~\bibnamefont {Ye}},\ }\href {\doibase
  10.1126/science.1184121} {\bibfield  {journal} {\bibinfo  {journal} {Science
  (80-. ).}\ }\textbf {\bibinfo {volume} {327}},\ \bibinfo {pages} {853}
  (\bibinfo {year} {2010})}\BibitemShut {NoStop}%
\bibitem [{\citenamefont {Hu}\ \emph {et~al.}(2019)\citenamefont {Hu},
  \citenamefont {Liu}, \citenamefont {Grimes}, \citenamefont {Lin},
  \citenamefont {Gheorghe}, \citenamefont {Vexiau}, \citenamefont
  {Bouloufa-Maafa}, \citenamefont {Dulieu}, \citenamefont {Rosenband},\ and\
  \citenamefont {Ni}}]{Hu2019}%
  \BibitemOpen
  \bibfield  {author} {\bibinfo {author} {\bibfnamefont {M.-G.}\ \bibnamefont
  {Hu}}, \bibinfo {author} {\bibfnamefont {Y.}~\bibnamefont {Liu}}, \bibinfo
  {author} {\bibfnamefont {D.~D.}\ \bibnamefont {Grimes}}, \bibinfo {author}
  {\bibfnamefont {Y.-W.}\ \bibnamefont {Lin}}, \bibinfo {author} {\bibfnamefont
  {A.~H.}\ \bibnamefont {Gheorghe}}, \bibinfo {author} {\bibfnamefont
  {R.}~\bibnamefont {Vexiau}}, \bibinfo {author} {\bibfnamefont
  {N.}~\bibnamefont {Bouloufa-Maafa}}, \bibinfo {author} {\bibfnamefont
  {O.}~\bibnamefont {Dulieu}}, \bibinfo {author} {\bibfnamefont
  {T.}~\bibnamefont {Rosenband}}, \ and\ \bibinfo {author} {\bibfnamefont
  {K.-K.}\ \bibnamefont {Ni}},\ }\href {\doibase 10.1126/science.aay9531}
  {\bibfield  {journal} {\bibinfo  {journal} {Science}\ }\textbf {\bibinfo
  {volume} {366}},\ \bibinfo {pages} {1111} (\bibinfo {year}
  {2019})}\BibitemShut {NoStop}%
\bibitem [{\citenamefont {Rabl}\ \emph {et~al.}(2006)\citenamefont {Rabl},
  \citenamefont {DeMille}, \citenamefont {Doyle}, \citenamefont {Lukin},
  \citenamefont {Schoelkopf},\ and\ \citenamefont {Zoller}}]{Rabl2006}%
  \BibitemOpen
  \bibfield  {author} {\bibinfo {author} {\bibfnamefont {P.}~\bibnamefont
  {Rabl}}, \bibinfo {author} {\bibfnamefont {D.}~\bibnamefont {DeMille}},
  \bibinfo {author} {\bibfnamefont {J.~M.}\ \bibnamefont {Doyle}}, \bibinfo
  {author} {\bibfnamefont {M.~D.}\ \bibnamefont {Lukin}}, \bibinfo {author}
  {\bibfnamefont {R.~J.}\ \bibnamefont {Schoelkopf}}, \ and\ \bibinfo {author}
  {\bibfnamefont {P.}~\bibnamefont {Zoller}},\ }\href {\doibase
  10.1103/PhysRevLett.97.033003} {\bibfield  {journal} {\bibinfo  {journal}
  {Phys. Rev. Lett.}\ }\textbf {\bibinfo {volume} {97}},\ \bibinfo {pages}
  {033003} (\bibinfo {year} {2006})}\BibitemShut {NoStop}%
\bibitem [{\citenamefont {DeMille}(2002)}]{DeMille2002}%
  \BibitemOpen
  \bibfield  {author} {\bibinfo {author} {\bibfnamefont {D.}~\bibnamefont
  {DeMille}},\ }\href {\doibase 10.1103/PhysRevLett.88.067901} {\bibfield
  {journal} {\bibinfo  {journal} {Phys. Rev. Lett.}\ }\textbf {\bibinfo
  {volume} {88}},\ \bibinfo {pages} {067901} (\bibinfo {year}
  {2002})}\BibitemShut {NoStop}%
\bibitem [{\citenamefont {Carr}\ \emph {et~al.}(2009)\citenamefont {Carr},
  \citenamefont {DeMille}, \citenamefont {Krems},\ and\ \citenamefont
  {Ye}}]{Carr2009}%
  \BibitemOpen
  \bibfield  {author} {\bibinfo {author} {\bibfnamefont {L.~D.}\ \bibnamefont
  {Carr}}, \bibinfo {author} {\bibfnamefont {D.}~\bibnamefont {DeMille}},
  \bibinfo {author} {\bibfnamefont {R.~V.}\ \bibnamefont {Krems}}, \ and\
  \bibinfo {author} {\bibfnamefont {J.}~\bibnamefont {Ye}},\ }\href {\doibase
  10.1088/1367-2630/11/5/055049} {\bibfield  {journal} {\bibinfo  {journal}
  {New Journal of Physics}\ }\textbf {\bibinfo {volume} {11}},\ \bibinfo
  {pages} {055049} (\bibinfo {year} {2009})}\BibitemShut {NoStop}%
\bibitem [{\citenamefont {Targo\ifmmode~\acute{n}\else \'{n}\fi{}ska}\ and\
  \citenamefont {Sacha}(2010)}]{Targonskatargonska2010}%
  \BibitemOpen
  \bibfield  {author} {\bibinfo {author} {\bibfnamefont {K.}~\bibnamefont
  {Targo\ifmmode~\acute{n}\else \'{n}\fi{}ska}}\ and\ \bibinfo {author}
  {\bibfnamefont {K.}~\bibnamefont {Sacha}},\ }\href {\doibase
  10.1103/PhysRevA.82.033601} {\bibfield  {journal} {\bibinfo  {journal} {Phys.
  Rev. A}\ }\textbf {\bibinfo {volume} {82}},\ \bibinfo {pages} {033601}
  (\bibinfo {year} {2010})}\BibitemShut {NoStop}%
\bibitem [{\citenamefont {Spiegelhalder}\ \emph {et~al.}(2009)\citenamefont
  {Spiegelhalder}, \citenamefont {Trenkwalder}, \citenamefont {Naik},
  \citenamefont {Hendl}, \citenamefont {Schreck},\ and\ \citenamefont
  {Grimm}}]{Spiegelhalder2009}%
  \BibitemOpen
  \bibfield  {author} {\bibinfo {author} {\bibfnamefont {F.~M.}\ \bibnamefont
  {Spiegelhalder}}, \bibinfo {author} {\bibfnamefont {A.}~\bibnamefont
  {Trenkwalder}}, \bibinfo {author} {\bibfnamefont {D.}~\bibnamefont {Naik}},
  \bibinfo {author} {\bibfnamefont {G.}~\bibnamefont {Hendl}}, \bibinfo
  {author} {\bibfnamefont {F.}~\bibnamefont {Schreck}}, \ and\ \bibinfo
  {author} {\bibfnamefont {R.}~\bibnamefont {Grimm}},\ }\href {\doibase
  10.1103/PhysRevLett.103.223203} {\bibfield  {journal} {\bibinfo  {journal}
  {Phys. Rev. Lett.}\ }\textbf {\bibinfo {volume} {103}},\ \bibinfo {pages}
  {223203} (\bibinfo {year} {2009})}\BibitemShut {NoStop}%
\bibitem [{\citenamefont {Gezerlis}\ \emph {et~al.}(2009)\citenamefont
  {Gezerlis}, \citenamefont {Gandolfi}, \citenamefont {Schmidt},\ and\
  \citenamefont {Carlson}}]{Gezerlis2009}%
  \BibitemOpen
  \bibfield  {author} {\bibinfo {author} {\bibfnamefont {A.}~\bibnamefont
  {Gezerlis}}, \bibinfo {author} {\bibfnamefont {S.}~\bibnamefont {Gandolfi}},
  \bibinfo {author} {\bibfnamefont {K.~E.}\ \bibnamefont {Schmidt}}, \ and\
  \bibinfo {author} {\bibfnamefont {J.}~\bibnamefont {Carlson}},\ }\href
  {\doibase 10.1103/PhysRevLett.103.060403} {\bibfield  {journal} {\bibinfo
  {journal} {Phys. Rev. Lett.}\ }\textbf {\bibinfo {volume} {103}},\ \bibinfo
  {pages} {060403} (\bibinfo {year} {2009})}\BibitemShut {NoStop}%
\bibitem [{\citenamefont {Trenkwalder}\ \emph {et~al.}(2011)\citenamefont
  {Trenkwalder}, \citenamefont {Kohstall}, \citenamefont {Zaccanti},
  \citenamefont {Naik}, \citenamefont {Sidorov}, \citenamefont {Schreck},\ and\
  \citenamefont {Grimm}}]{Trenkwalder2011}%
  \BibitemOpen
  \bibfield  {author} {\bibinfo {author} {\bibfnamefont {A.}~\bibnamefont
  {Trenkwalder}}, \bibinfo {author} {\bibfnamefont {C.}~\bibnamefont
  {Kohstall}}, \bibinfo {author} {\bibfnamefont {M.}~\bibnamefont {Zaccanti}},
  \bibinfo {author} {\bibfnamefont {D.}~\bibnamefont {Naik}}, \bibinfo {author}
  {\bibfnamefont {A.~I.}\ \bibnamefont {Sidorov}}, \bibinfo {author}
  {\bibfnamefont {F.}~\bibnamefont {Schreck}}, \ and\ \bibinfo {author}
  {\bibfnamefont {R.}~\bibnamefont {Grimm}},\ }\href {\doibase
  10.1103/PhysRevLett.106.115304} {\bibfield  {journal} {\bibinfo  {journal}
  {Phys. Rev. Lett.}\ }\textbf {\bibinfo {volume} {106}},\ \bibinfo {pages}
  {115304} (\bibinfo {year} {2011})}\BibitemShut {NoStop}%
\bibitem [{\citenamefont {Wu}\ \emph {et~al.}(2006)\citenamefont {Wu},
  \citenamefont {Pao},\ and\ \citenamefont {Yip}}]{Wu2006}%
  \BibitemOpen
  \bibfield  {author} {\bibinfo {author} {\bibfnamefont {S.-T.}\ \bibnamefont
  {Wu}}, \bibinfo {author} {\bibfnamefont {C.-H.}\ \bibnamefont {Pao}}, \ and\
  \bibinfo {author} {\bibfnamefont {S.-K.}\ \bibnamefont {Yip}},\ }\href
  {\doibase 10.1103/PhysRevB.74.224504} {\bibfield  {journal} {\bibinfo
  {journal} {Phys. Rev. B}\ }\textbf {\bibinfo {volume} {74}},\ \bibinfo
  {pages} {224504} (\bibinfo {year} {2006})}\BibitemShut {NoStop}%
\bibitem [{\citenamefont {Iskin}(2008)}]{Iskin2008}%
  \BibitemOpen
  \bibfield  {author} {\bibinfo {author} {\bibfnamefont {M.}~\bibnamefont
  {Iskin}},\ }\href {\doibase 10.1103/PhysRevA.78.021604} {\bibfield  {journal}
  {\bibinfo  {journal} {Phys. Rev. A}\ }\textbf {\bibinfo {volume} {78}},\
  \bibinfo {pages} {021604} (\bibinfo {year} {2008})}\BibitemShut {NoStop}%
\bibitem [{\citenamefont {Pires}\ \emph {et~al.}(2014)\citenamefont {Pires},
  \citenamefont {Ulmanis}, \citenamefont {H\"afner}, \citenamefont {Repp},
  \citenamefont {Arias}, \citenamefont {Kuhnle},\ and\ \citenamefont
  {Weidem\"uller}}]{Pires2014}%
  \BibitemOpen
  \bibfield  {author} {\bibinfo {author} {\bibfnamefont {R.}~\bibnamefont
  {Pires}}, \bibinfo {author} {\bibfnamefont {J.}~\bibnamefont {Ulmanis}},
  \bibinfo {author} {\bibfnamefont {S.}~\bibnamefont {H\"afner}}, \bibinfo
  {author} {\bibfnamefont {M.}~\bibnamefont {Repp}}, \bibinfo {author}
  {\bibfnamefont {A.}~\bibnamefont {Arias}}, \bibinfo {author} {\bibfnamefont
  {E.~D.}\ \bibnamefont {Kuhnle}}, \ and\ \bibinfo {author} {\bibfnamefont
  {M.}~\bibnamefont {Weidem\"uller}},\ }\href {\doibase
  10.1103/PhysRevLett.112.250404} {\bibfield  {journal} {\bibinfo  {journal}
  {Phys. Rev. Lett.}\ }\textbf {\bibinfo {volume} {112}},\ \bibinfo {pages}
  {250404} (\bibinfo {year} {2014})}\BibitemShut {NoStop}%
\bibitem [{\citenamefont {D'Incao}\ and\ \citenamefont
  {Esry}(2006)}]{DIncao2006}%
  \BibitemOpen
  \bibfield  {author} {\bibinfo {author} {\bibfnamefont {J.~P.}\ \bibnamefont
  {D'Incao}}\ and\ \bibinfo {author} {\bibfnamefont {B.~D.}\ \bibnamefont
  {Esry}},\ }\href {\doibase 10.1103/PhysRevA.73.030702} {\bibfield  {journal}
  {\bibinfo  {journal} {Phys. Rev. A}\ }\textbf {\bibinfo {volume} {73}},\
  \bibinfo {pages} {030702} (\bibinfo {year} {2006})}\BibitemShut {NoStop}%
\bibitem [{\citenamefont {Gu\'erout}\ \emph {et~al.}(2010)\citenamefont
  {Gu\'erout}, \citenamefont {Aymar},\ and\ \citenamefont
  {Dulieu}}]{Dulieu2010SrLipotential}%
  \BibitemOpen
  \bibfield  {author} {\bibinfo {author} {\bibfnamefont {R.}~\bibnamefont
  {Gu\'erout}}, \bibinfo {author} {\bibfnamefont {M.}~\bibnamefont {Aymar}}, \
  and\ \bibinfo {author} {\bibfnamefont {O.}~\bibnamefont {Dulieu}},\ }\href
  {\doibase 10.1103/PhysRevA.82.042508} {\bibfield  {journal} {\bibinfo
  {journal} {Phys. Rev. A}\ }\textbf {\bibinfo {volume} {82}},\ \bibinfo
  {pages} {042508} (\bibinfo {year} {2010})}\BibitemShut {NoStop}%
\bibitem [{\citenamefont {Gopakumar}\ \emph {et~al.}(2013)\citenamefont
  {Gopakumar}, \citenamefont {Abe}, \citenamefont {Hada},\ and\ \citenamefont
  {Kajita}}]{gopakumar2013ab}%
  \BibitemOpen
  \bibfield  {author} {\bibinfo {author} {\bibfnamefont {G.}~\bibnamefont
  {Gopakumar}}, \bibinfo {author} {\bibfnamefont {M.}~\bibnamefont {Abe}},
  \bibinfo {author} {\bibfnamefont {M.}~\bibnamefont {Hada}}, \ and\ \bibinfo
  {author} {\bibfnamefont {M.}~\bibnamefont {Kajita}},\ }\href {\doibase
  10.1063/1.4804622} {\bibfield  {journal} {\bibinfo  {journal} {The Journal of
  Chemical Physics}\ }\textbf {\bibinfo {volume} {138}},\ \bibinfo {pages}
  {194307} (\bibinfo {year} {2013})}\BibitemShut {NoStop}%
\bibitem [{\citenamefont {Mitroy}\ and\ \citenamefont
  {Bromley}(2003)}]{Mitroy2003}%
  \BibitemOpen
  \bibfield  {author} {\bibinfo {author} {\bibfnamefont {J.}~\bibnamefont
  {Mitroy}}\ and\ \bibinfo {author} {\bibfnamefont {M.~W.~J.}\ \bibnamefont
  {Bromley}},\ }\href {\doibase 10.1103/PhysRevA.68.052714} {\bibfield
  {journal} {\bibinfo  {journal} {Phys. Rev. A}\ }\textbf {\bibinfo {volume}
  {68}},\ \bibinfo {pages} {052714} (\bibinfo {year} {2003})}\BibitemShut
  {NoStop}%
\bibitem [{\citenamefont {Standard}\ and\ \citenamefont
  {Certain}(1985)}]{Standard1985}%
  \BibitemOpen
  \bibfield  {author} {\bibinfo {author} {\bibfnamefont {J.~M.}\ \bibnamefont
  {Standard}}\ and\ \bibinfo {author} {\bibfnamefont {P.~R.}\ \bibnamefont
  {Certain}},\ }\href {\doibase 10.1063/1.449203} {\bibfield  {journal}
  {\bibinfo  {journal} {The Journal of Chemical Physics}\ }\textbf {\bibinfo
  {volume} {83}},\ \bibinfo {pages} {3002} (\bibinfo {year}
  {1985})}\BibitemShut {NoStop}%
\bibitem [{\citenamefont {Stanton}(1994)}]{Stanton1994}%
  \BibitemOpen
  \bibfield  {author} {\bibinfo {author} {\bibfnamefont {J.~F.}\ \bibnamefont
  {Stanton}},\ }\href {\doibase 10.1103/PhysRevA.49.1698} {\bibfield  {journal}
  {\bibinfo  {journal} {Phys. Rev. A}\ }\textbf {\bibinfo {volume} {49}},\
  \bibinfo {pages} {1698} (\bibinfo {year} {1994})}\BibitemShut {NoStop}%
\bibitem [{\citenamefont {Kotochigova}\ \emph {et~al.}(2011)\citenamefont
  {Kotochigova}, \citenamefont {Petrov}, \citenamefont {Linnik}, \citenamefont
  {K?os},\ and\ \citenamefont {Julienne}}]{Kotochigova2011}%
  \BibitemOpen
  \bibfield  {author} {\bibinfo {author} {\bibfnamefont {S.}~\bibnamefont
  {Kotochigova}}, \bibinfo {author} {\bibfnamefont {A.}~\bibnamefont {Petrov}},
  \bibinfo {author} {\bibfnamefont {M.}~\bibnamefont {Linnik}}, \bibinfo
  {author} {\bibfnamefont {J.}~\bibnamefont {K?os}}, \ and\ \bibinfo {author}
  {\bibfnamefont {P.~S.}\ \bibnamefont {Julienne}},\ }\href {\doibase
  10.1063/1.3653974} {\bibfield  {journal} {\bibinfo  {journal} {The Journal of
  Chemical Physics}\ }\textbf {\bibinfo {volume} {135}},\ \bibinfo {pages}
  {164108} (\bibinfo {year} {2011})}\BibitemShut {NoStop}%
\bibitem [{\citenamefont {Chen}\ \emph {et~al.}(2014)\citenamefont {Chen},
  \citenamefont {Zhu}, \citenamefont {Li}, \citenamefont {Qian},\ and\
  \citenamefont {Wang}}]{Chen2014}%
  \BibitemOpen
  \bibfield  {author} {\bibinfo {author} {\bibfnamefont {T.}~\bibnamefont
  {Chen}}, \bibinfo {author} {\bibfnamefont {S.}~\bibnamefont {Zhu}}, \bibinfo
  {author} {\bibfnamefont {X.}~\bibnamefont {Li}}, \bibinfo {author}
  {\bibfnamefont {J.}~\bibnamefont {Qian}}, \ and\ \bibinfo {author}
  {\bibfnamefont {Y.}~\bibnamefont {Wang}},\ }\href {\doibase
  10.1103/PhysRevA.89.063402} {\bibfield  {journal} {\bibinfo  {journal} {Phys.
  Rev. A}\ }\textbf {\bibinfo {volume} {89}},\ \bibinfo {pages} {063402}
  (\bibinfo {year} {2014})}\BibitemShut {NoStop}%
\bibitem [{\citenamefont {Pototschnig}\ \emph {et~al.}(2016)\citenamefont
  {Pototschnig}, \citenamefont {Hauser},\ and\ \citenamefont
  {Ernst}}]{Pototschnig2016}%
  \BibitemOpen
  \bibfield  {author} {\bibinfo {author} {\bibfnamefont {J.~V.}\ \bibnamefont
  {Pototschnig}}, \bibinfo {author} {\bibfnamefont {A.~W.}\ \bibnamefont
  {Hauser}}, \ and\ \bibinfo {author} {\bibfnamefont {W.~E.}\ \bibnamefont
  {Ernst}},\ }\href {\doibase 10.1039/c5cp06598d} {\bibfield  {journal}
  {\bibinfo  {journal} {Physical Chemistry Chemical Physics}\ }\textbf
  {\bibinfo {volume} {18}},\ \bibinfo {pages} {5964} (\bibinfo {year}
  {2016})}\BibitemShut {NoStop}%
\bibitem [{\citenamefont {Augustovi{\v{c}}ov{\'a}}\ and\ \citenamefont
  {Sold{\'a}n}(2012)}]{Augustovicova2012}%
  \BibitemOpen
  \bibfield  {author} {\bibinfo {author} {\bibfnamefont {L.}~\bibnamefont
  {Augustovi{\v{c}}ov{\'a}}}\ and\ \bibinfo {author} {\bibfnamefont
  {P.}~\bibnamefont {Sold{\'a}n}},\ }\href {\doibase 10.1063/1.3690459}
  {\bibfield  {journal} {\bibinfo  {journal} {The Journal of Chemical Physics}\
  }\textbf {\bibinfo {volume} {136}},\ \bibinfo {pages} {084311} (\bibinfo
  {year} {2012})}\BibitemShut {NoStop}%
\bibitem [{\citenamefont {Tomza}(2013)}]{Tomza2013}%
  \BibitemOpen
  \bibfield  {author} {\bibinfo {author} {\bibfnamefont {M.}~\bibnamefont
  {Tomza}},\ }\href {\doibase 10.1103/PhysRevA.88.012519} {\bibfield  {journal}
  {\bibinfo  {journal} {Phys. Rev. A}\ }\textbf {\bibinfo {volume} {88}},\
  \bibinfo {pages} {012519} (\bibinfo {year} {2013})}\BibitemShut {NoStop}%
\bibitem [{\citenamefont {Guo}\ \emph {et~al.}(2013)\citenamefont {Guo},
  \citenamefont {Bajdich}, \citenamefont {Mitas},\ and\ \citenamefont
  {Reynolds}}]{Guo2013}%
  \BibitemOpen
  \bibfield  {author} {\bibinfo {author} {\bibfnamefont {S.}~\bibnamefont
  {Guo}}, \bibinfo {author} {\bibfnamefont {M.}~\bibnamefont {Bajdich}},
  \bibinfo {author} {\bibfnamefont {L.}~\bibnamefont {Mitas}}, \ and\ \bibinfo
  {author} {\bibfnamefont {P.~J.}\ \bibnamefont {Reynolds}},\ }\href {\doibase
  10.1080/00268976.2013.788741} {\bibfield  {journal} {\bibinfo  {journal}
  {Molecular Physics}\ }\textbf {\bibinfo {volume} {111}},\ \bibinfo {pages}
  {1744} (\bibinfo {year} {2013})}\BibitemShut {NoStop}%
\bibitem [{\citenamefont {Shao}\ \emph {et~al.}(2017)\citenamefont {Shao},
  \citenamefont {Deng}, \citenamefont {Xing}, \citenamefont {Gou},
  \citenamefont {Kuang},\ and\ \citenamefont {Li}}]{shao2017ground}%
  \BibitemOpen
  \bibfield  {author} {\bibinfo {author} {\bibfnamefont {Q.}~\bibnamefont
  {Shao}}, \bibinfo {author} {\bibfnamefont {L.}~\bibnamefont {Deng}}, \bibinfo
  {author} {\bibfnamefont {X.}~\bibnamefont {Xing}}, \bibinfo {author}
  {\bibfnamefont {D.}~\bibnamefont {Gou}}, \bibinfo {author} {\bibfnamefont
  {X.}~\bibnamefont {Kuang}}, \ and\ \bibinfo {author} {\bibfnamefont
  {H.}~\bibnamefont {Li}},\ }\href
  {https://pubs.acs.org/doi/abs/10.1021/acs.jpca.6b11741} {\bibfield  {journal}
  {\bibinfo  {journal} {The Journal of Physical Chemistry A}\ }\textbf
  {\bibinfo {volume} {121}},\ \bibinfo {pages} {2187} (\bibinfo {year}
  {2017})}\BibitemShut {NoStop}%
\bibitem [{\citenamefont {Tohme}\ and\ \citenamefont
  {Korek}(2015)}]{Tohme2015}%
  \BibitemOpen
  \bibfield  {author} {\bibinfo {author} {\bibfnamefont {S.~N.}\ \bibnamefont
  {Tohme}}\ and\ \bibinfo {author} {\bibfnamefont {M.}~\bibnamefont {Korek}},\
  }\href {\doibase https://doi.org/10.1016/j.cplett.2015.08.050} {\bibfield
  {journal} {\bibinfo  {journal} {Chemical Physics Letters}\ }\textbf {\bibinfo
  {volume} {638}},\ \bibinfo {pages} {216 } (\bibinfo {year}
  {2015})}\BibitemShut {NoStop}%
\bibitem [{\citenamefont {Allouche}\ and\ \citenamefont
  {Aubert-Fr¨¨con}(1994)}]{Allouche1994}%
  \BibitemOpen
  \bibfield  {author} {\bibinfo {author} {\bibfnamefont {A.}~\bibnamefont
  {Allouche}}\ and\ \bibinfo {author} {\bibfnamefont {M.}~\bibnamefont
  {Aubert-Fr¨¨con}},\ }\href {\doibase
  https://doi.org/10.1016/0009-2614(94)00371-8} {\bibfield  {journal} {\bibinfo
   {journal} {Chemical Physics Letters}\ }\textbf {\bibinfo {volume} {222}},\
  \bibinfo {pages} {524 } (\bibinfo {year} {1994})}\BibitemShut {NoStop}%
\bibitem [{\citenamefont {Kajita}\ \emph {et~al.}(2012)\citenamefont {Kajita},
  \citenamefont {Gopakumar}, \citenamefont {Abe},\ and\ \citenamefont
  {Hada}}]{kajita2012}%
  \BibitemOpen
  \bibfield  {author} {\bibinfo {author} {\bibfnamefont {M.}~\bibnamefont
  {Kajita}}, \bibinfo {author} {\bibfnamefont {G.}~\bibnamefont {Gopakumar}},
  \bibinfo {author} {\bibfnamefont {M.}~\bibnamefont {Abe}}, \ and\ \bibinfo
  {author} {\bibfnamefont {M.}~\bibnamefont {Hada}},\ }\href
  {https://iopscience.iop.org/article/10.1088/0953-4075/46/2/025001/meta}
  {\bibfield  {journal} {\bibinfo  {journal} {Journal of Physics B: Atomic,
  Molecular and Optical Physics}\ }\textbf {\bibinfo {volume} {46}},\ \bibinfo
  {pages} {025001} (\bibinfo {year} {2012})}\BibitemShut {NoStop}%
\bibitem [{\citenamefont {Brue}\ and\ \citenamefont {Hutson}(2012)}]{Brue2012}%
  \BibitemOpen
  \bibfield  {author} {\bibinfo {author} {\bibfnamefont {D.~A.}\ \bibnamefont
  {Brue}}\ and\ \bibinfo {author} {\bibfnamefont {J.~M.}\ \bibnamefont
  {Hutson}},\ }\href {\doibase 10.1103/PhysRevLett.108.043201} {\bibfield
  {journal} {\bibinfo  {journal} {Phys. Rev. Lett.}\ }\textbf {\bibinfo
  {volume} {108}},\ \bibinfo {pages} {043201} (\bibinfo {year}
  {2012})}\BibitemShut {NoStop}%
\bibitem [{\citenamefont {Zeid}\ \emph {et~al.}(2018)\citenamefont {Zeid},
  \citenamefont {Atallah}, \citenamefont {Kontar}, \citenamefont {Chmaisani},
  \citenamefont {El-Kork},\ and\ \citenamefont {Korek}}]{MahmoudKoreka2018SrX}%
  \BibitemOpen
  \bibfield  {author} {\bibinfo {author} {\bibfnamefont {I.}~\bibnamefont
  {Zeid}}, \bibinfo {author} {\bibfnamefont {T.}~\bibnamefont {Atallah}},
  \bibinfo {author} {\bibfnamefont {S.}~\bibnamefont {Kontar}}, \bibinfo
  {author} {\bibfnamefont {W.}~\bibnamefont {Chmaisani}}, \bibinfo {author}
  {\bibfnamefont {N.}~\bibnamefont {El-Kork}}, \ and\ \bibinfo {author}
  {\bibfnamefont {M.}~\bibnamefont {Korek}},\ }\href {\doibase
  https://doi.org/10.1016/j.comptc.2018.01.013} {\bibfield  {journal} {\bibinfo
   {journal} {Computational and Theoretical Chemistry}\ }\textbf {\bibinfo
  {volume} {1126}},\ \bibinfo {pages} {16 } (\bibinfo {year}
  {2018})}\BibitemShut {NoStop}%
\bibitem [{\citenamefont {Hara}\ \emph {et~al.}(2011)\citenamefont {Hara},
  \citenamefont {Takasu}, \citenamefont {Yamaoka}, \citenamefont {Doyle},\ and\
  \citenamefont {Takahashi}}]{Hara2011}%
  \BibitemOpen
  \bibfield  {author} {\bibinfo {author} {\bibfnamefont {H.}~\bibnamefont
  {Hara}}, \bibinfo {author} {\bibfnamefont {Y.}~\bibnamefont {Takasu}},
  \bibinfo {author} {\bibfnamefont {Y.}~\bibnamefont {Yamaoka}}, \bibinfo
  {author} {\bibfnamefont {J.~M.}\ \bibnamefont {Doyle}}, \ and\ \bibinfo
  {author} {\bibfnamefont {Y.}~\bibnamefont {Takahashi}},\ }\href {\doibase
  10.1103/PhysRevLett.106.205304} {\bibfield  {journal} {\bibinfo  {journal}
  {Phys. Rev. Lett.}\ }\textbf {\bibinfo {volume} {106}},\ \bibinfo {pages}
  {205304} (\bibinfo {year} {2011})}\BibitemShut {NoStop}%
\bibitem [{\citenamefont {Hansen}\ \emph {et~al.}(2011)\citenamefont {Hansen},
  \citenamefont {Khramov}, \citenamefont {Dowd}, \citenamefont {Jamison},
  \citenamefont {Ivanov},\ and\ \citenamefont {Gupta}}]{Hansen2011}%
  \BibitemOpen
  \bibfield  {author} {\bibinfo {author} {\bibfnamefont {A.~H.}\ \bibnamefont
  {Hansen}}, \bibinfo {author} {\bibfnamefont {A.}~\bibnamefont {Khramov}},
  \bibinfo {author} {\bibfnamefont {W.~H.}\ \bibnamefont {Dowd}}, \bibinfo
  {author} {\bibfnamefont {A.~O.}\ \bibnamefont {Jamison}}, \bibinfo {author}
  {\bibfnamefont {V.~V.}\ \bibnamefont {Ivanov}}, \ and\ \bibinfo {author}
  {\bibfnamefont {S.}~\bibnamefont {Gupta}},\ }\href {\doibase
  10.1103/PhysRevA.84.011606} {\bibfield  {journal} {\bibinfo  {journal} {Phys.
  Rev. A}\ }\textbf {\bibinfo {volume} {84}},\ \bibinfo {pages} {011606}
  (\bibinfo {year} {2011})}\BibitemShut {NoStop}%
\bibitem [{\citenamefont {Vaidya}\ \emph {et~al.}(2015)\citenamefont {Vaidya},
  \citenamefont {Tiamsuphat}, \citenamefont {Rolston},\ and\ \citenamefont
  {Porto}}]{Vaidya2015}%
  \BibitemOpen
  \bibfield  {author} {\bibinfo {author} {\bibfnamefont {V.~D.}\ \bibnamefont
  {Vaidya}}, \bibinfo {author} {\bibfnamefont {J.}~\bibnamefont {Tiamsuphat}},
  \bibinfo {author} {\bibfnamefont {S.~L.}\ \bibnamefont {Rolston}}, \ and\
  \bibinfo {author} {\bibfnamefont {J.~V.}\ \bibnamefont {Porto}},\ }\href
  {\doibase 10.1103/PhysRevA.92.043604} {\bibfield  {journal} {\bibinfo
  {journal} {Phys. Rev. A}\ }\textbf {\bibinfo {volume} {92}},\ \bibinfo
  {pages} {043604} (\bibinfo {year} {2015})}\BibitemShut {NoStop}%
\bibitem [{\citenamefont {Pasquiou}\ \emph {et~al.}(2013)\citenamefont
  {Pasquiou}, \citenamefont {Bayerle}, \citenamefont {Tzanova}, \citenamefont
  {Stellmer}, \citenamefont {Szczepkowski}, \citenamefont {Parigger},
  \citenamefont {Grimm},\ and\ \citenamefont {Schreck}}]{Pasquiou2013}%
  \BibitemOpen
  \bibfield  {author} {\bibinfo {author} {\bibfnamefont {B.}~\bibnamefont
  {Pasquiou}}, \bibinfo {author} {\bibfnamefont {A.}~\bibnamefont {Bayerle}},
  \bibinfo {author} {\bibfnamefont {S.~M.}\ \bibnamefont {Tzanova}}, \bibinfo
  {author} {\bibfnamefont {S.}~\bibnamefont {Stellmer}}, \bibinfo {author}
  {\bibfnamefont {J.}~\bibnamefont {Szczepkowski}}, \bibinfo {author}
  {\bibfnamefont {M.}~\bibnamefont {Parigger}}, \bibinfo {author}
  {\bibfnamefont {R.}~\bibnamefont {Grimm}}, \ and\ \bibinfo {author}
  {\bibfnamefont {F.}~\bibnamefont {Schreck}},\ }\href {\doibase
  10.1103/PhysRevA.88.023601} {\bibfield  {journal} {\bibinfo  {journal} {Phys.
  Rev. A}\ }\textbf {\bibinfo {volume} {88}},\ \bibinfo {pages} {023601}
  (\bibinfo {year} {2013})}\BibitemShut {NoStop}%
\bibitem [{\citenamefont {Guttridge}\ \emph {et~al.}(2018)\citenamefont
  {Guttridge}, \citenamefont {Hopkins}, \citenamefont {Frye}, \citenamefont
  {McFerran}, \citenamefont {Hutson},\ and\ \citenamefont
  {Cornish}}]{Guttridge2018}%
  \BibitemOpen
  \bibfield  {author} {\bibinfo {author} {\bibfnamefont {A.}~\bibnamefont
  {Guttridge}}, \bibinfo {author} {\bibfnamefont {S.~A.}\ \bibnamefont
  {Hopkins}}, \bibinfo {author} {\bibfnamefont {M.~D.}\ \bibnamefont {Frye}},
  \bibinfo {author} {\bibfnamefont {J.~J.}\ \bibnamefont {McFerran}}, \bibinfo
  {author} {\bibfnamefont {J.~M.}\ \bibnamefont {Hutson}}, \ and\ \bibinfo
  {author} {\bibfnamefont {S.~L.}\ \bibnamefont {Cornish}},\ }\href {\doibase
  10.1103/PhysRevA.97.063414} {\bibfield  {journal} {\bibinfo  {journal} {Phys.
  Rev. A}\ }\textbf {\bibinfo {volume} {97}},\ \bibinfo {pages} {063414}
  (\bibinfo {year} {2018})}\BibitemShut {NoStop}%
\bibitem [{\citenamefont {Aoki}\ \emph {et~al.}(2013)\citenamefont {Aoki},
  \citenamefont {Yamanaka}, \citenamefont {Takeuchi}, \citenamefont {Torii},\
  and\ \citenamefont {Sakemi}}]{Aoki2013}%
  \BibitemOpen
  \bibfield  {author} {\bibinfo {author} {\bibfnamefont {T.}~\bibnamefont
  {Aoki}}, \bibinfo {author} {\bibfnamefont {Y.}~\bibnamefont {Yamanaka}},
  \bibinfo {author} {\bibfnamefont {M.}~\bibnamefont {Takeuchi}}, \bibinfo
  {author} {\bibfnamefont {Y.}~\bibnamefont {Torii}}, \ and\ \bibinfo {author}
  {\bibfnamefont {Y.}~\bibnamefont {Sakemi}},\ }\href {\doibase
  10.1103/PhysRevA.87.063426} {\bibfield  {journal} {\bibinfo  {journal} {Phys.
  Rev. A}\ }\textbf {\bibinfo {volume} {87}},\ \bibinfo {pages} {063426}
  (\bibinfo {year} {2013})}\BibitemShut {NoStop}%
\bibitem [{\citenamefont {Bruni}\ and\ \citenamefont
  {G\"orlitz}(2016)}]{Bruni2016}%
  \BibitemOpen
  \bibfield  {author} {\bibinfo {author} {\bibfnamefont {C.}~\bibnamefont
  {Bruni}}\ and\ \bibinfo {author} {\bibfnamefont {A.}~\bibnamefont
  {G\"orlitz}},\ }\href {\doibase 10.1103/PhysRevA.94.022503} {\bibfield
  {journal} {\bibinfo  {journal} {Phys. Rev. A}\ }\textbf {\bibinfo {volume}
  {94}},\ \bibinfo {pages} {022503} (\bibinfo {year} {2016})}\BibitemShut
  {NoStop}%
\bibitem [{\citenamefont {Schwanke}\ \emph {et~al.}(2017)\citenamefont
  {Schwanke}, \citenamefont {Kn\"{o}ckel}, \citenamefont {Stein}, \citenamefont
  {Pashov}, \citenamefont {Ospelkaus},\ and\ \citenamefont
  {Tiemann}}]{Schwanke2017}%
  \BibitemOpen
  \bibfield  {author} {\bibinfo {author} {\bibfnamefont {E.}~\bibnamefont
  {Schwanke}}, \bibinfo {author} {\bibfnamefont {H.}~\bibnamefont
  {Kn\"{o}ckel}}, \bibinfo {author} {\bibfnamefont {A.}~\bibnamefont {Stein}},
  \bibinfo {author} {\bibfnamefont {A.}~\bibnamefont {Pashov}}, \bibinfo
  {author} {\bibfnamefont {S.}~\bibnamefont {Ospelkaus}}, \ and\ \bibinfo
  {author} {\bibfnamefont {E.}~\bibnamefont {Tiemann}},\ }\href {\doibase
  10.1088/1361-6455/aa8ca0} {\bibfield  {journal} {\bibinfo  {journal} {Journal
  of Physics B: Atomic, Molecular and Optical Physics}\ }\textbf {\bibinfo
  {volume} {50}},\ \bibinfo {pages} {235103} (\bibinfo {year}
  {2017})}\BibitemShut {NoStop}%
\bibitem [{\citenamefont {Kraft}\ \emph {et~al.}(2009)\citenamefont {Kraft},
  \citenamefont {Vogt}, \citenamefont {Appel}, \citenamefont {Riehle},\ and\
  \citenamefont {Sterr}}]{kraft2009Ca}%
  \BibitemOpen
  \bibfield  {author} {\bibinfo {author} {\bibfnamefont {S.}~\bibnamefont
  {Kraft}}, \bibinfo {author} {\bibfnamefont {F.}~\bibnamefont {Vogt}},
  \bibinfo {author} {\bibfnamefont {O.}~\bibnamefont {Appel}}, \bibinfo
  {author} {\bibfnamefont {F.}~\bibnamefont {Riehle}}, \ and\ \bibinfo {author}
  {\bibfnamefont {U.}~\bibnamefont {Sterr}},\ }\href {\doibase
  10.1103/PhysRevLett.103.130401} {\bibfield  {journal} {\bibinfo  {journal}
  {Physical review letters}\ }\textbf {\bibinfo {volume} {103}},\ \bibinfo
  {pages} {130401} (\bibinfo {year} {2009})}\BibitemShut {NoStop}%
\bibitem [{\citenamefont {Lackner}\ \emph {et~al.}(2014)\citenamefont
  {Lackner}, \citenamefont {Krois}, \citenamefont {Buchsteiner}, \citenamefont
  {Pototschnig},\ and\ \citenamefont {Ernst}}]{WolfgangE2014DroplrtRbSr}%
  \BibitemOpen
  \bibfield  {author} {\bibinfo {author} {\bibfnamefont {F.}~\bibnamefont
  {Lackner}}, \bibinfo {author} {\bibfnamefont {G.}~\bibnamefont {Krois}},
  \bibinfo {author} {\bibfnamefont {T.}~\bibnamefont {Buchsteiner}}, \bibinfo
  {author} {\bibfnamefont {J.~V.}\ \bibnamefont {Pototschnig}}, \ and\ \bibinfo
  {author} {\bibfnamefont {W.~E.}\ \bibnamefont {Ernst}},\ }\href {\doibase
  10.1103/PhysRevLett.113.153001} {\bibfield  {journal} {\bibinfo  {journal}
  {Phys. Rev. Lett.}\ }\textbf {\bibinfo {volume} {113}},\ \bibinfo {pages}
  {153001} (\bibinfo {year} {2014})}\BibitemShut {NoStop}%
\bibitem [{\citenamefont {Russon}\ \emph {et~al.}(1998)\citenamefont {Russon},
  \citenamefont {Rothschopf}, \citenamefont {Morse}, \citenamefont {Boldyrev},\
  and\ \citenamefont {Simons}}]{JackSimons1998LiCa}%
  \BibitemOpen
  \bibfield  {author} {\bibinfo {author} {\bibfnamefont {L.~M.}\ \bibnamefont
  {Russon}}, \bibinfo {author} {\bibfnamefont {G.~K.}\ \bibnamefont
  {Rothschopf}}, \bibinfo {author} {\bibfnamefont {M.~D.}\ \bibnamefont
  {Morse}}, \bibinfo {author} {\bibfnamefont {A.~I.}\ \bibnamefont {Boldyrev}},
  \ and\ \bibinfo {author} {\bibfnamefont {J.}~\bibnamefont {Simons}},\ }\href
  {\doibase 10.1063/1.477317} {\bibfield  {journal} {\bibinfo  {journal} {The
  Journal of Chemical Physics}\ }\textbf {\bibinfo {volume} {109}},\ \bibinfo
  {pages} {6655} (\bibinfo {year} {1998})},\ \Eprint
  {http://arxiv.org/abs/https://doi.org/10.1063/1.477317}
  {https://doi.org/10.1063/1.477317} \BibitemShut {NoStop}%
\bibitem [{\citenamefont {D¡¯Incan}\ \emph {et~al.}(1994)\citenamefont
  {D¡¯Incan}, \citenamefont {Effantin}, \citenamefont {Bernard}, \citenamefont
  {Fabre}, \citenamefont {Stringat}, \citenamefont {Boulezhar},\ and\
  \citenamefont {Verg¨¨s}}]{JVerges1994BaLi}%
  \BibitemOpen
  \bibfield  {author} {\bibinfo {author} {\bibfnamefont {J.}~\bibnamefont
  {D¡¯Incan}}, \bibinfo {author} {\bibfnamefont {C.}~\bibnamefont {Effantin}},
  \bibinfo {author} {\bibfnamefont {A.}~\bibnamefont {Bernard}}, \bibinfo
  {author} {\bibfnamefont {G.}~\bibnamefont {Fabre}}, \bibinfo {author}
  {\bibfnamefont {R.}~\bibnamefont {Stringat}}, \bibinfo {author}
  {\bibfnamefont {A.}~\bibnamefont {Boulezhar}}, \ and\ \bibinfo {author}
  {\bibfnamefont {J.}~\bibnamefont {Verg¨¨s}},\ }\href {\doibase
  10.1063/1.466576} {\bibfield  {journal} {\bibinfo  {journal} {The Journal of
  Chemical Physics}\ }\textbf {\bibinfo {volume} {100}},\ \bibinfo {pages}
  {945} (\bibinfo {year} {1994})},\ \Eprint
  {http://arxiv.org/abs/https://doi.org/10.1063/1.466576}
  {https://doi.org/10.1063/1.466576} \BibitemShut {NoStop}%
\bibitem [{\citenamefont {Barb{\'e}}\ \emph {et~al.}(2018)\citenamefont
  {Barb{\'e}}, \citenamefont {Ciamei}, \citenamefont {Pasquiou}, \citenamefont
  {Reichs{\"o}llner}, \citenamefont {Schreck}, \citenamefont
  {{\.{Z}}uchowski},\ and\ \citenamefont {Hutson}}]{Barbe2018}%
  \BibitemOpen
  \bibfield  {author} {\bibinfo {author} {\bibfnamefont {V.}~\bibnamefont
  {Barb{\'e}}}, \bibinfo {author} {\bibfnamefont {A.}~\bibnamefont {Ciamei}},
  \bibinfo {author} {\bibfnamefont {B.}~\bibnamefont {Pasquiou}}, \bibinfo
  {author} {\bibfnamefont {L.}~\bibnamefont {Reichs{\"o}llner}}, \bibinfo
  {author} {\bibfnamefont {F.}~\bibnamefont {Schreck}}, \bibinfo {author}
  {\bibfnamefont {P.~S.}\ \bibnamefont {{\.{Z}}uchowski}}, \ and\ \bibinfo
  {author} {\bibfnamefont {J.~M.}\ \bibnamefont {Hutson}},\ }\href {\doibase
  10.1038/s41567-018-0169-x} {\bibfield  {journal} {\bibinfo  {journal} {Nature
  Physics}\ }\textbf {\bibinfo {volume} {14}},\ \bibinfo {pages} {881}
  (\bibinfo {year} {2018})}\BibitemShut {NoStop}%
\bibitem [{\citenamefont {Green}\ \emph {et~al.}(2019)\citenamefont {Green},
  \citenamefont {Li}, \citenamefont {Toh}, \citenamefont {Tang}, \citenamefont
  {McCormick}, \citenamefont {Li}, \citenamefont {Tiesinga}, \citenamefont
  {Kotochigova},\ and\ \citenamefont {Gupta}}]{green2019feshbach}%
  \BibitemOpen
  \bibfield  {author} {\bibinfo {author} {\bibfnamefont {A.}~\bibnamefont
  {Green}}, \bibinfo {author} {\bibfnamefont {H.}~\bibnamefont {Li}}, \bibinfo
  {author} {\bibfnamefont {J.~H.~S.}\ \bibnamefont {Toh}}, \bibinfo {author}
  {\bibfnamefont {X.}~\bibnamefont {Tang}}, \bibinfo {author} {\bibfnamefont
  {K.}~\bibnamefont {McCormick}}, \bibinfo {author} {\bibfnamefont
  {M.}~\bibnamefont {Li}}, \bibinfo {author} {\bibfnamefont {E.}~\bibnamefont
  {Tiesinga}}, \bibinfo {author} {\bibfnamefont {S.}~\bibnamefont
  {Kotochigova}}, \ and\ \bibinfo {author} {\bibfnamefont {S.}~\bibnamefont
  {Gupta}},\ }\href {https://arxiv.org/abs/1912.04874} {\bibfield  {journal}
  {\bibinfo  {journal} {arXiv preprint arXiv:1912.04874}\ } (\bibinfo {year}
  {2019})}\BibitemShut {NoStop}%
\bibitem [{\citenamefont {Micheli}\ \emph {et~al.}(2006)\citenamefont
  {Micheli}, \citenamefont {Brennen},\ and\ \citenamefont
  {Zoller}}]{Micheli2006}%
  \BibitemOpen
  \bibfield  {author} {\bibinfo {author} {\bibfnamefont {A.}~\bibnamefont
  {Micheli}}, \bibinfo {author} {\bibfnamefont {G.~K.}\ \bibnamefont
  {Brennen}}, \ and\ \bibinfo {author} {\bibfnamefont {P.}~\bibnamefont
  {Zoller}},\ }\href {\doibase 10.1038/nphys287} {\bibfield  {journal}
  {\bibinfo  {journal} {Nat. Phys.}\ }\textbf {\bibinfo {volume} {2}},\
  \bibinfo {pages} {341} (\bibinfo {year} {2006})}\BibitemShut {NoStop}%
\bibitem [{\citenamefont {P{\'e}rez-R{\'\i}os}\ \emph
  {et~al.}(2010)\citenamefont {P{\'e}rez-R{\'\i}os}, \citenamefont {Herrera},\
  and\ \citenamefont {Krems}}]{perez2010}%
  \BibitemOpen
  \bibfield  {author} {\bibinfo {author} {\bibfnamefont {J.}~\bibnamefont
  {P{\'e}rez-R{\'\i}os}}, \bibinfo {author} {\bibfnamefont {F.}~\bibnamefont
  {Herrera}}, \ and\ \bibinfo {author} {\bibfnamefont {R.~V.}\ \bibnamefont
  {Krems}},\ }\href
  {https://iopscience.iop.org/article/10.1088/1367-2630/12/10/103007}
  {\bibfield  {journal} {\bibinfo  {journal} {New Journal of Physics}\ }\textbf
  {\bibinfo {volume} {12}},\ \bibinfo {pages} {103007} (\bibinfo {year}
  {2010})}\BibitemShut {NoStop}%
\bibitem [{\citenamefont {Meyer}\ and\ \citenamefont {Bohn}(2009)}]{Meyer2009}%
  \BibitemOpen
  \bibfield  {author} {\bibinfo {author} {\bibfnamefont {E.~R.}\ \bibnamefont
  {Meyer}}\ and\ \bibinfo {author} {\bibfnamefont {J.~L.}\ \bibnamefont
  {Bohn}},\ }\href {\doibase 10.1103/PhysRevA.80.042508} {\bibfield  {journal}
  {\bibinfo  {journal} {Phys. Rev. A}\ }\textbf {\bibinfo {volume} {80}},\
  \bibinfo {pages} {042508} (\bibinfo {year} {2009})}\BibitemShut {NoStop}%
\bibitem [{\citenamefont {Ma}\ \emph {et~al.}(2019)\citenamefont {Ma},
  \citenamefont {Ye}, \citenamefont {Xie}, \citenamefont {Guo}, \citenamefont
  {You},\ and\ \citenamefont {Tey}}]{Ma2019}%
  \BibitemOpen
  \bibfield  {author} {\bibinfo {author} {\bibfnamefont {X.-B.}\ \bibnamefont
  {Ma}}, \bibinfo {author} {\bibfnamefont {Z.-X.}\ \bibnamefont {Ye}}, \bibinfo
  {author} {\bibfnamefont {L.-Y.}\ \bibnamefont {Xie}}, \bibinfo {author}
  {\bibfnamefont {Z.}~\bibnamefont {Guo}}, \bibinfo {author} {\bibfnamefont
  {L.}~\bibnamefont {You}}, \ and\ \bibinfo {author} {\bibfnamefont {M.~K.}\
  \bibnamefont {Tey}},\ }\href {\doibase 10.1088/0256-307x/36/7/073401}
  {\bibfield  {journal} {\bibinfo  {journal} {Chinese Phys. Lett.}\ }\textbf
  {\bibinfo {volume} {36}},\ \bibinfo {pages} {073401} (\bibinfo {year}
  {2019})}\BibitemShut {NoStop}%
\bibitem [{Ma2(2019)}]{Ma2019Erratum}%
  \BibitemOpen
  \href {\doibase 10.1088/0256-307x/36/10/109902} {\bibfield  {journal}
  {\bibinfo  {journal} {Chinese Physics Letters}\ }\textbf {\bibinfo {volume}
  {36}},\ \bibinfo {pages} {109902} (\bibinfo {year} {2019})}\BibitemShut
  {NoStop}%
\bibitem [{\citenamefont {Stellmer}\ \emph {et~al.}(2009)\citenamefont
  {Stellmer}, \citenamefont {Tey}, \citenamefont {Huang}, \citenamefont
  {Grimm},\ and\ \citenamefont {Schreck}}]{Stellmer2009}%
  \BibitemOpen
  \bibfield  {author} {\bibinfo {author} {\bibfnamefont {S.}~\bibnamefont
  {Stellmer}}, \bibinfo {author} {\bibfnamefont {M.~K.}\ \bibnamefont {Tey}},
  \bibinfo {author} {\bibfnamefont {B.}~\bibnamefont {Huang}}, \bibinfo
  {author} {\bibfnamefont {R.}~\bibnamefont {Grimm}}, \ and\ \bibinfo {author}
  {\bibfnamefont {F.}~\bibnamefont {Schreck}},\ }\href {\doibase
  10.1103/PhysRevLett.103.200401} {\bibfield  {journal} {\bibinfo  {journal}
  {Phys. Rev. Lett.}\ }\textbf {\bibinfo {volume} {103}},\ \bibinfo {pages}
  {200401} (\bibinfo {year} {2009})}\BibitemShut {NoStop}%
\bibitem [{\citenamefont {de~Escobar}\ \emph {et~al.}(2009)\citenamefont
  {de~Escobar}, \citenamefont {Mickelson}, \citenamefont {Yan}, \citenamefont
  {DeSalvo}, \citenamefont {Nagel},\ and\ \citenamefont
  {Killian}}]{Killian2009SrBEC}%
  \BibitemOpen
  \bibfield  {author} {\bibinfo {author} {\bibfnamefont {Y.~N.~M.}\
  \bibnamefont {de~Escobar}}, \bibinfo {author} {\bibfnamefont {P.~G.}\
  \bibnamefont {Mickelson}}, \bibinfo {author} {\bibfnamefont {M.}~\bibnamefont
  {Yan}}, \bibinfo {author} {\bibfnamefont {B.~J.}\ \bibnamefont {DeSalvo}},
  \bibinfo {author} {\bibfnamefont {S.~B.}\ \bibnamefont {Nagel}}, \ and\
  \bibinfo {author} {\bibfnamefont {T.~C.}\ \bibnamefont {Killian}},\ }\href
  {\doibase 10.1103/PhysRevLett.103.200402} {\bibfield  {journal} {\bibinfo
  {journal} {Phys. Rev. Lett.}\ }\textbf {\bibinfo {volume} {103}},\ \bibinfo
  {pages} {200402} (\bibinfo {year} {2009})}\BibitemShut {NoStop}%
\bibitem [{coo()}]{coolingbeams}%
  \BibitemOpen
  \href@noop {} {}\bibinfo {note} {The overlapped cooling light beams include
  671-nm light (with 4 different frequencies) for realizing MOT and gray
  molasses of $^{6}$Li, and 461-nm and 671-nm light for realizing blue MOT and
  red MOT of $^{84}$Sr.}\BibitemShut {Stop}%
\bibitem [{dip()}]{dipoletraplasers}%
  \BibitemOpen
  \href@noop {} {}\bibinfo {note} {The polarizations of the horizontal beams
  are both linear but made orthogonal to each other to avoid interference. The
  41$^\circ$-tilted beam, which is derived from a different laser, interferes
  with the horizontal ones but at a beating frequency too fast to cause
  heating.}\BibitemShut {Stop}%
\bibitem [{\citenamefont {Burchianti}\ \emph {et~al.}(2014)\citenamefont
  {Burchianti}, \citenamefont {Valtolina}, \citenamefont {Seman}, \citenamefont
  {Pace}, \citenamefont {De~Pas}, \citenamefont {Inguscio}, \citenamefont
  {Zaccanti},\ and\ \citenamefont {Roati}}]{Burchianti2014}%
  \BibitemOpen
  \bibfield  {author} {\bibinfo {author} {\bibfnamefont {A.}~\bibnamefont
  {Burchianti}}, \bibinfo {author} {\bibfnamefont {G.}~\bibnamefont
  {Valtolina}}, \bibinfo {author} {\bibfnamefont {J.~A.}\ \bibnamefont
  {Seman}}, \bibinfo {author} {\bibfnamefont {E.}~\bibnamefont {Pace}},
  \bibinfo {author} {\bibfnamefont {M.}~\bibnamefont {De~Pas}}, \bibinfo
  {author} {\bibfnamefont {M.}~\bibnamefont {Inguscio}}, \bibinfo {author}
  {\bibfnamefont {M.}~\bibnamefont {Zaccanti}}, \ and\ \bibinfo {author}
  {\bibfnamefont {G.}~\bibnamefont {Roati}},\ }\href {\doibase
  10.1103/PhysRevA.90.043408} {\bibfield  {journal} {\bibinfo  {journal} {Phys.
  Rev. A}\ }\textbf {\bibinfo {volume} {90}},\ \bibinfo {pages} {043408}
  (\bibinfo {year} {2014})}\BibitemShut {NoStop}%
\bibitem [{\citenamefont {Gallagher}\ and\ \citenamefont
  {Pritchard}(1989)}]{Gallagher1989}%
  \BibitemOpen
  \bibfield  {author} {\bibinfo {author} {\bibfnamefont {A.}~\bibnamefont
  {Gallagher}}\ and\ \bibinfo {author} {\bibfnamefont {D.~E.}\ \bibnamefont
  {Pritchard}},\ }\href {\doibase 10.1103/PhysRevLett.63.957} {\bibfield
  {journal} {\bibinfo  {journal} {Phys. Rev. Lett.}\ }\textbf {\bibinfo
  {volume} {63}},\ \bibinfo {pages} {957} (\bibinfo {year} {1989})}\BibitemShut
  {NoStop}%
\bibitem [{\citenamefont {O'Hara}\ \emph {et~al.}(2001)\citenamefont {O'Hara},
  \citenamefont {Gehm}, \citenamefont {Granade},\ and\ \citenamefont
  {Thomas}}]{OHara2001}%
  \BibitemOpen
  \bibfield  {author} {\bibinfo {author} {\bibfnamefont {K.~M.}\ \bibnamefont
  {O'Hara}}, \bibinfo {author} {\bibfnamefont {M.~E.}\ \bibnamefont {Gehm}},
  \bibinfo {author} {\bibfnamefont {S.~R.}\ \bibnamefont {Granade}}, \ and\
  \bibinfo {author} {\bibfnamefont {J.~E.}\ \bibnamefont {Thomas}},\ }\href
  {\doibase 10.1103/PhysRevA.64.051403} {\bibfield  {journal} {\bibinfo
  {journal} {Phys. Rev. A}\ }\textbf {\bibinfo {volume} {64}},\ \bibinfo
  {pages} {051403} (\bibinfo {year} {2001})}\BibitemShut {NoStop}%
\bibitem [{int()}]{integratedFermiDirac}%
  \BibitemOpen
  \href@noop {} {}\bibinfo {note} {This profile is a good approximation for a
  spatially integrated Fermi-Dirac distribution for a gas initially trapped in
  a harmonic potential and after a time of flight larger than the trap
  oscillation periods.}\BibitemShut {Stop}%
\bibitem [{\citenamefont {Tey}\ \emph {et~al.}(2010)\citenamefont {Tey},
  \citenamefont {Stellmer}, \citenamefont {Grimm},\ and\ \citenamefont
  {Schreck}}]{Tey2010doubledegenerate}%
  \BibitemOpen
  \bibfield  {author} {\bibinfo {author} {\bibfnamefont {M.~K.}\ \bibnamefont
  {Tey}}, \bibinfo {author} {\bibfnamefont {S.}~\bibnamefont {Stellmer}},
  \bibinfo {author} {\bibfnamefont {R.}~\bibnamefont {Grimm}}, \ and\ \bibinfo
  {author} {\bibfnamefont {F.}~\bibnamefont {Schreck}},\ }\href {\doibase
  10.1103/PhysRevA.82.011608} {\bibfield  {journal} {\bibinfo  {journal} {Phys.
  Rev. A}\ }\textbf {\bibinfo {volume} {82}},\ \bibinfo {pages} {011608}
  (\bibinfo {year} {2010})}\BibitemShut {NoStop}%
\bibitem [{the()}]{thermalizationTimeConst}%
  \BibitemOpen
  \href@noop {} {}\bibinfo {note} {We calculate the thermalization time
  constant $\tau$ using equation \begin{equation}\label{eq:thermalrate}
  \tau=-\frac{1}{\Delta T}\frac{d(\Delta
  T)}{dt}=\frac{\xi}{\alpha}\bar{n}\sigma \bar{v}. \end{equation} Here, $\Delta
  T=T_1-T_2$ is the temperature difference between thermal component 1 and 2.
  $\alpha=2.7$ is the average times of collisions needed for thermalization
  between colliding pairs with equal mass~\cite{Wieman1993CsCs}, and
  $\xi=\frac{4 m_1 m_2}{(m_1+m_2)^2}$ is a correction factor to $\alpha$ for
  nonequal mass partners ($\xi=0.249$ for $^{6}$Li-$^{84}$Sr mixture). The mean
  relative velocity
  $\bar{v}=\sqrt{\frac{8k_B}{\pi}(\frac{T_1}{m_1}+\frac{T_2}{m_2})}$, and the
  overlap density $\bar{n}=\left(\frac{1}{N_1}+\frac{1}{N_2}\right)\int n_1(
  \bm{r})n_2(\bm{r})d^3r$ with \begin{equation}\label{eq:density} \int
  n_1(\bm{r})n_2(\bm{r})d^3r=\frac{N_1 N_2 \left(m_1 m_2 \bar{\omega}^2_1
  \bar{\omega}^2_2\right)^{3/2}}{\left[2\pi k_B(m_1\bar{\omega}^2_1
  T_2+m_2\bar{\omega}^2_2 T_1)\right]^{3/2}}, \end{equation} with $N_1$($N_2$)
  being the number of atoms of component 1(2).}\BibitemShut {Stop}%
\bibitem [{\citenamefont {Monroe}\ \emph {et~al.}(1993)\citenamefont {Monroe},
  \citenamefont {Cornell}, \citenamefont {Sackett}, \citenamefont {Myatt},\
  and\ \citenamefont {Wieman}}]{Wieman1993CsCs}%
  \BibitemOpen
  \bibfield  {author} {\bibinfo {author} {\bibfnamefont {C.~R.}\ \bibnamefont
  {Monroe}}, \bibinfo {author} {\bibfnamefont {E.~A.}\ \bibnamefont {Cornell}},
  \bibinfo {author} {\bibfnamefont {C.~A.}\ \bibnamefont {Sackett}}, \bibinfo
  {author} {\bibfnamefont {C.~J.}\ \bibnamefont {Myatt}}, \ and\ \bibinfo
  {author} {\bibfnamefont {C.~E.}\ \bibnamefont {Wieman}},\ }\href {\doibase
  10.1103/PhysRevLett.70.414} {\bibfield  {journal} {\bibinfo  {journal} {Phys.
  Rev. Lett.}\ }\textbf {\bibinfo {volume} {70}},\ \bibinfo {pages} {414}
  (\bibinfo {year} {1993})}\BibitemShut {NoStop}%
\end{thebibliography}

%

\end{document}